\documentclass[english]{emulateapj}
\usepackage{subfigure}

\usepackage[T1]{fontenc}
\usepackage[latin1]{inputenc}
\usepackage{graphicx}
\usepackage{amssymb}

\makeatletter

\usepackage{times}
\usepackage{amsmath}

\newcommand{\peryr}{\,{\rm yr^{-1}}}

\newcommand{\kmpers}{\,{\rm km s^{-1}}}
\newcommand{\yr}{{\,\rm yr}}
\newcommand{\vdisp}{v_{\rm disp}}

\newcommand{\Gyr}{{\,\rm Gyr}}

\newcommand{\Myr}{{\,\rm Myr}}
\newcommand{\pc}{\,\mathrm{pc}}

\newcommand{\Mbh}{M_{\bullet}}
\newcommand{\Ro}{R_{\odot}}
\newcommand{\Mo}{M_{\odot}}

\newcommand{\AU}{{\,\rm AU}}

\shorttitle{BINARY DYNAMICS NEAR A MASSIVE BLACK HOLE}
\shortauthors{HOPMAN}

\usepackage{babel}
\makeatother
\begin{document}
\bibliographystyle{apj.bst} 
%

%
%

\title{Binary dynamics near a massive black hole}

\author{Clovis Hopman\footnote{Leiden University, Leiden Observatory, P.O. Box 9513,
NL-2300 RA Leiden}}

\affiliation{Leiden University, Leiden Observatory, P.O. Box 9513, NL-2300 RA Leiden}

\email{clovis@strw.leidenuniv.nl}

\begin{abstract}
We analyze the dynamical evolution of binary stars that interact with a static background of 
single stars in the environment of a massive black hole (MBH). All
stars are considered to be single mass, Newtonian point particles. We
follow the evolution of the energy $E$ and angular momentum $J$ of the
center of mass of the binaries with respect to the MBH, as well as
their internal semi-major axis $a$, using a Monte Carlo method.

For a system like the Galactic center, the main conclusions are the
following: (1) The binary fraction can be of the order of a few
percent outside 0.1 pc, but decreases quickly closer to the
MBH. (2) Within $\sim$0.1 pc, binaries can only exist on eccentric orbits with apocenters 
much further away from the MBH. (3) Far away from the
MBH, loss-cone effects are the dominant mechanism that disrupts
binaries with internal velocities close to the velocity
dispersion. Closer to the MBH, three-body encounters are more effective in
disrupting binaries. (4) The rate at which
hard binaries become tighter is usually less than the rate at which a binary diffuses
to orbits that are more bound to the MBH. (5) Binaries are typically
disrupted before they experience an exchange interaction; as a result,
the number of exchanges is less than one would estimate from a simple
``$nv\sigma$ estimate''. 

We give applications of our results to the formation of X-ray binaries near MBHs and to the production rates of hyper-velocity stars by intermediate mass MBHs.
\end{abstract}

\keywords{black hole physics  --- stellar dynamics --- Galaxy: center --- binaries: general}

\section{Introduction}\label{s:intro}
The presence of a massive black hole (MBH) in a stellar system has
several interesting implications for stellar dynamics in those
systems \citep[see e.g.][]{Ale03c}. Black holes provide the deepest gravitational potential wells
in nature. As a result, MBHs give rise to a strong tidal field, and extremely high velocity
dispersions which diverge as $\vdisp\propto \sqrt{1/r}$. At the same time, the stellar density in these systems also diverges towards the center. In spite of the high velocities, the rate of encounters between stars can therefore be sufficiently high for the system to exhibit relaxational evolution within the age of the Universe. In this paper we explore the dynamical evolution of binary stars under the conditions of a very hot, dense system with a tidal sink.

\subsection{Collisional stellar dynamics near a massive black hole}\label{s:reldyn}

The relaxation time is for a single mass system conventionally defined
as

\begin{equation}\label{e:tr}
t_r\equiv {0.34\vdisp^3\over (Gm)^2 n \ln\Lambda},
\end{equation}
where $m$ is the mass of a star, $n$ is the number density, $\vdisp$ is the velocity dispersion, and
$\ln\Lambda$ is the Coulomb logarithm \citep{Spi71, Bin08}. This time
scale is a measure for the time it takes for a system to evolve as a result of the sum of many weak, uncorrelated two-body interactions. The resulting density profile of single mass stars close to a MBH was
first found by \citet[][]{Bah76}. They solved the
one-dimensional (only energy) Fokker-Planck equations under the
approximation that the potential of the system is given by
the MBH. In steady state, the stellar profile is given by

\begin{equation}\label{e:BW}
n(r)\propto r^{-7/4};\quad N(E)\propto E^{-9/4},
\end{equation}
where $r$ is the radial distance from the MBH, $E=G\Mbh/r-v^2/2$ the negative specific energy with respect to the MBH for a star with velocity $v$ (hereafter "energy"); and $N(E)dE$ is the number of stars in the interval\footnote{We will use the symbol $N$ for several functions, with the argument implying its meaning.} $(E,
E+dE)$. Extensive theoretical work has confirmed the existence of a
\citet{Bah76} profile under a wide variety of assumptions and methods,
\citep[e.g. ][]{Sha78, Coh78, Fre02, Bau04a, Pre04}, including $N$-body simulations with primordial binaries \citep{Tre07}. There is now also
observational support of a diverging density profile that is approximately
consistent with \cite{Bah76} in the Galactic center \citep{Ale99a, Gen03a,
Sch07}. It is important to note, however, that the interpretation of
relaxational stellar dynamics in the Galactic center is complicated by
the presence of many luminous young stars, which dominate the light,
but are certainly not relaxed \citep[e.g.][]{Lev03, Pau06, Bar08}, and giants, which may be affected by hydrodynamical collisions \citep{Ale99a}. For a comprehensive review of stellar
phenomena and observations thereof in the Galactic center, see
\citet{Ale05}.

From equations (\ref{e:tr}, \ref{e:BW}), it can be seen that for a
\citet{Bah76} profile, the relaxation time decreases as $r^{1/4}$ as $r$
decreases, in spite of the diverging velocity dispersion. Many systems, perhaps most notably galaxies as a whole, have relaxation times that
are many orders of magnitude larger than the age of the Universe, and
the effect of the diffusion of stars as a result of two-body
interactions is entirely negligible. However, close to MBHs of masses
$\Mbh \lesssim 10^7 \Mo$, the relaxation time is smaller than the age
of the Universe. Such systems are known as collisional, and give rise to a
variety of phenomena that are absent for systems with much larger
relaxation times. These phenomena include the formation of the \citet{Bah76}
cusp; mass segregation \citep[e.g.][]{Bah77, Mur91, Fre06, Hop06b, OLe08, Ale08, Kes09}; loss-cone
dynamics \citep[e.g.][for more details see below]{Fra76, Lig77, Bah77,
Coh78, Sye99, Mag99, Ama04}; inspiral of stars due to gravitational waves or tidal
dissipation \citep[e.g.][]{Hil95, Fre01, Fre03, Hop04, Hop05, Bau05b};
and evolution of binaries, which is the subject of this paper.

In summary, the dynamics in the close vicinity of a MBH is
characterized by very high velocity dispersions; very high densities;
and short relaxation times. These three aspects lead to an extreme
environment for stellar processes.

\subsection{Binaries}

Binary stars play a major role in many branches of astronomy. Interest
in binaries from a dynamical perspective has greatly increased
starting with the seminal papers by \citet{Heg75} and
\citet{Hil75}. From a dynamical perspective binaries have a dual role.

When (single) stars interact with binaries, the internal properties of
the binaries change. The binary can become tighter or less tight, it
can dissolve, and it can exchange one of its members with an incoming
star. It is therefore interesting to study the evolution of the
binaries themselves as they interact with their environment.

In some cases, binaries can also have a dominating role on the
environment around them. In the point particle approximation, a binary
can be infinitely tight, and hence provide an infinite source of
energy. As a result, binaries appear to play a major role in stopping
core-collapse \citep[e.g.][]{Goo89, Gao91, Hut92}.

In the environment of a MBH that is of interest here, binaries do not
dominate the energetics, since stars can become very bound to the MBH, and
hence the energies of the single stars also become naturally very
large. This conclusion is confirmed by $N$-body simulations, which show that binaries do not affect the global evolution of a stellar system with a MBH \citep{Tre07}. It is interesting, however, to consider how they react to their
environment. Interactions with single stars and with the MBH can
modify their distribution and composition.

When all stars are of single mass $m$, which is what is assumed
throughout this paper, there is a natural dimensionless number

\begin{equation}\label{e:hs}
\xi\equiv {\epsilon\over m \vdisp^2} = {G m\over 2a\vdisp^2}
\end{equation}
that characterizes a binary with semi-major axis $a$ moving through a
sea of stars that is characterized by a velocity dispersion $v_{\rm disp}$. We use the symbol $\epsilon=Gm^2/(2a)$ for the negative internal energy of the binary (hereafter "energy"; note that in contrast to the energy $E$ with respect to the MBH, this energy is not specific).

Binaries with $\xi\ll1$ are called ``soft''. It is established by
extensive theoretical and numerical work that on average, such
binaries will become even softer, so that $\xi\to0$, until they
eventually dissolve \citep[e.g.][]{Heg75, Hut83, Hut83b}. Binaries with $\xi\gg1$ are called ``hard''. In interactions with their environment, such binaries tend to become even harder
\citep[e.g.][]{Heg75, Hut93a, Heg96}.

From a dynamical perspective, the evolution of binaries near a MBH is
interesting for several reasons. 

As outlined above, the close environment near a MBH is unique in that
it has a diverging density profile, a diverging velocity field, and a
sufficiently short relaxation time to allow for dynamical evolution. Typically, the rate at which binaries interact
with single stars is of the order of the relaxation rate divided by
the Coulomb logarithm (see \S \ref{s:times} for a more quantitative
comparison). A short relaxation time therefore implies a high
single-binary interaction rate. The fact that the velocity dispersion
diverges as the radius from the MBH $r$ decreases, implies that a
binary that is hard at large $r$, may well be soft at smaller
$r$. As this binary diffuses inwards due to two-body interactions, it therefore first hardens due
to interactions with the cold stars around it. But if it does not
do so at a sufficiently high rate, it will at smaller $r$ be
surrounded by very hot stars, soften and ionize at smaller $r$.

Another aspect that is unique for the system considered in this paper,
is that at small $r$ the tidal field from the MBH can be much larger
than the forces that bind the binary components; tidal disruptions
therefore form another channel of binary destruction (\S
\ref{s:destruct}).

While the distribution of binary stars is an interesting theoretical
problem in stellar dynamics, it also has a number of direct
observational implications. We will not investigate all of these phenomena in this paper, but list them as a motivation of the interest in binary dynamics near MBHs.

{\it X-ray binaries --- } Within the inner $\pc$ of the Galactic
center, there is an over-abundance of X-ray binaries compared to the
bulge \citep{Mun05}. It has long been known that globular clusters,
which also have small relaxation times, contain X-ray binaries in
numbers per unit mass that are orders of magnitude larger than the
rest of the bulge \citep[see e.g.][for a review]{Heg03}, and it is
thought that collisional dynamics plays a decisive role in this
overabundance \citep{Poo03}. In this paper the number of stars that
has experienced an exchange interaction is followed, and from this the
number of compact remnants present in binaries is estimated. The results are discussed in \S\ref{s:disc}.

{\it Ultra-luminous X-ray sources ---} When massive binaries are tidally disrupted by an intermediate mass black hole (IMBH) of $\Mbh=10^3-10^4\Mo$, the captured star may after expansion feed the black hole. This possibility was proposed as a possible explanation for ultra-luminous X-ray sources \citep{Ble06}.

{\it Pulsars --- } Tidal disruption of binaries with compact remnants may lead to the presence of neutron stars, and perhaps millisecond radio pulsars very close to IMBHs. Detection of such a pulsar so close to an IMBH would give strong evidence for the presence of the IMBH \citep{Pfa04, Pfa05, Pat05}

{\it Hyper velocity stars --- } If a binary is disrupted in the tidal
field of a MBH, one of the components can be ejected at a very large
speed ($\sim 10^3 \kmpers$) from the Galaxy \citep{Hil88, YuQ03}.  
These hyper-velocity stars are now observed \citep{Bro05, Brown06, Bro07, Bro08a}. In \S\ref{s:disc} we comment on the possibility of ejection of hyper-velocity stars from an intermediate mass black hole in the Large Magellanic Cloud.

{\it S-stars --- } The energy carried by a hyper-velocity star
escaping from the energy is drawn from the other component of the
original binary, which ends up on a close orbit to the MBH. It has
been proposed that the cluster of young, randomly oriented stars in the Galactic center, known as the S-stars originated in this process
\citep{Gou03, Per07, Lok08, Mad08, Per09}.

{\it Gravitational waves: LISA --- } If one of the components of a
binary is a compact remnant, and it is on a tight orbit around the MBH after the
binary is tidally disrupted, it may later spiral in to the MBH while
emitting gravitational waves. This process may be of considerable
importance for gravitational wave detection of extreme mass ratio
inspiral sources by the {\it Laser Interferometer Space Antenna}
(LISA) \citep{Mil05}.

{\it Gravitational waves: LIGO --- } It was shown by \citet{OLe08}
that due to the extreme stellar densities, occasionally stellar black
hole binaries form in the close vicinity of the MBH. If such binaries
avoid subsequent tidal disruption and ionization, and they spiral in
within the age of the age of the Universe, they may be an important
source of short wave length gravitational waves that can be observed
by advanced LIGO.

{\it Blue stragglers --- } Mergers of the two components in a binary
star may lead to rejuvenation of a binary. Such rejuvenated stars are
known as ``blue stragglers'' in globular clusters; see \citet{Heg03}
for a review. This phenomena may also be of relevance in the formation of some of the hypervelocity stars \citep{Per07b}.

\subsection{Method}\label{s:method}

Several different techniques have been used to study stellar dynamics
in the vicinity of MBHs. The conceptually most straightforward method
is to integrate the system by direct $N$-body simulations \citep[e.g.,
][]{Bau04a, Bau04b, Pre04}. A study including primordial
binaries was made by \citet{Tre07}. The main disadvantage of direct
$N$-body simulations is that they are limited to an unrealistically
small number of particles; the largest study by \citet{Tre07} had
16384 particles, whereas a minimum of several millions of particles is
required to model the GC. In a situation where collisional dynamics is
driving the evolution, the outcome may depend crucially on the
particle number, so that straightforward scaling of the results is not always possible. In addition, the finite radius of the stars provides a new length scale; as a result, the binary distribution and disruption rate near $\Mbh=10^3\Mo$ IMBHs and $\Mbh=3\times10^6\Mo$ MBHs is quite different.

Other methods that have been used to study stellar dynamics near MBHs
rely explicitly or implicitly on the Fokker-Planck
approximation. Statistical approaches like solving Boltzmann equations
or Fokker-Planck equations (mostly using a Monte Carlo [MC] approach) have
also been used frequently in the past to find the evolution of the
semi-major axis of binaries \citep[e.g.][]{Gao91, Goo93, Ivanova05, Ble06}

Explicit integration of the Fokker-Planck equations in one dimension
(just the energies of the stars with respect to the MBH) were made first in the pioneering
work of \citet{Bah76, Bah77}, and later by several other authors
\citep[e.g. ][]{Mur91, Pre04, Hop06, Hop06b, OLe08, Ale08, Kes09}. There has
also been a two dimensional study by \citet{Coh78} that followed both the energies and
angular momenta of the stars with respect to the MBH using direct
integration. Direct integration of the Fokker-Planck
equations, along with gaseous models \citep{Ama04} is probably the fastest method to study the collisional
effects of interest here. The disadvantage of these methods is that it
is hard to implement numerically in an efficient way, and that it is
harder to adapt to include additional physics than MC
methods.

Fokker-Planck equations can also be solved by a MC method. In
this method, the distribution function (DF) is represented by a realization with an arbitrary number of particles, that evolves by a
combination of deterministic (drift) and random steps (diffusion). In recent years,
extensive use of the \citet{Hen73} implementation of this method has been
made \citep[e.g. ][]{Fre01b, Fre02, Gur04b, Fre06}. These
papers describe this method in much detail, and also compare it to
other methods. Here, we use a different,
simpler model here that is based on \citet{Sha78} \cite[see][for a discussion of the differences between the methods]{Fre01b}. This model is
easier to implement, it is suitable for the problem at hand, and it
keeps a structure that is closer to the original Fokker-Planck
equation. Due to easily implemented individual time steps, it also allows an accurate treatment of stars on very eccentric orbits, that could become tidally disrupted. The \citet{Sha78} method, and the current implementation, is
explained in some detail in \S\ref{s:MC}. We extend the two-dimensional (energy and angular momentum) \citet{Sha78} model to three dimensions, the third dimension being the internal semi-major axis of the binary.

\subsection{Plan}\label{s:plan}
This paper is organized as follows. In \S\ref{s:scaling} several
results about single mass stellar systems near MBHs are recapitulated,
and systems with different MBH masses are related through the relation between MBH mass and velocity dispersion of the host galaxy. Some of the more important assumptions used in
this paper are summarized. In \S\ref{ss:EJ} we discuss how we will
treat the evolution of the center of mass of the binaries with respect
to the MBH as these diffuse through energy and angular momentum
space. In \S\ref{s:bin} the evolution of the internal semi-major axis
of the binaries is described; the case of destruction of binaries is treated separately in \S \ref{s:destruct}. The dynamics in
\S\S\ref{ss:EJ} - \ref{s:destruct} give rise to several time-scales;
in \S\ref{s:times} these time-scales are compared, leading to a
qualitative overview of the dynamics, including a parametrized analytical model for the binary disruption rates. In \S\ref{s:MC} we describe the
MC scheme, in which the dynamics is implemented. In
\S\ref{s:results} a number of different cases are studied in detail
and the results are presented. In \S\ref{s:disc} the paper is
summarized and its results are discussed. We apply our results to the formation rate of X-ray binaries in the Galactic center, and to the possibility of the ejection of hyper-velocity stars by the Large Magellanic Cloud.

\section{Scaling relations of a single mass cusp}\label{s:scaling}
Throughout this paper, it is assumed that binary stars interact only
with a static background of single stars, and with a MBH. All stars are of Solar type. An important
simplification that directly follows from this is that binary-binary
interactions are neglected, and as a result that the evolution is
linear, i.e., the diffusion coefficients in the Fokker-Planck equation are independent of the DF of the binaries. Because of the high velocity dispersion near MBHs, only very
tight binaries can survive. The binary fraction near MBHs, even for
marginally bound orbits, is therefore expected to be of order of only a
few percent. This justifies neglecting mutual binary interactions to
first order.

The distribution of single stars of equal mass $m$ around a MBH of mass $\Mbh$ was first found by
\citet{Bah76}, who derived and solved the Fokker-Planck equations describing their
evolution in energy space. The shape of the resulting distribution
within the radius of influence of the MBH is independent of any of the
parameters (although it is of course subject to a number of
assumptions, such as that $\Mbh\gg m$). Here we recapitulate some of
their results. To scale systems with different MBH masses, the relation between the mass of MBHs, $\Mbh$, and the
velocity dispersion $v_0$ of their host bulge or galaxy
\citep{Fer00, Geb00, Tre02} is used,

\begin{equation}\label{e:Msigma}
M_{\bullet}=1.3\times10^{8}M_{\odot}\left(\frac{v_0}{200\,\mathrm{km\,
 s^{-1}}}\right)^{4}\,.
\end{equation}
Both the exponent and the pre-factor have small
uncertainties, which are ignored here. Although the
relation (\ref{e:Msigma}) has only been observed for masses $\Mbh\!\gtrsim\!10^6\Mo$, we
assume that it can be extrapolated to lower mass MBHs and IMBHs.

The relation (\ref{e:Msigma}) can then be exploited to scale several
quantities with MBH mass. The velocity dispersion for unbound
orbits is

\begin{equation}\label{e:sig}
v_{0} = 78 \kmpers \left({\Mbh\over3\times10^6\Mo}\right)^{1/4}.
\end{equation}

Unbound stars that come closer than a distance

\begin{equation}\label{e:rhMsig}
r_{0} =  {\frac{G\Mbh}{v_0^{2}}} = 2.1\,\mathrm{pc}\,\left({\frac{\Mbh}{3\times10^{6}\Mo}}\right)^{1/2}
\end{equation}
are strongly influenced by the MBH; $r_{0}$ is known as the
radius of influence. Our analysis will be restricted to distances less
than $r_{0}$. The amount of mass in stars within the radius of
influence is usually comparable to the MBH mass. This implies that the
density number density $n_{0}$ can be scaled with the observed density
in the Galactic center,

\begin{equation}\label{e:n0}
n_{0} = 4.5\times10^4\pc^{-3} \left({\Mbh\over3\times10^6\Mo}\right)^{-1/2}
\end{equation}
\citep{Gen03a}, where it was assumed that stars are of Solar mass.

The period at the radius of influence is
\begin{eqnarray}\label{e:P0}
P_{0} &=& 2\pi\left({r_h^3\over G\Mbh}\right)^{1/2} \nonumber\\
&=& 1.7\times10^{5}\yr \left({\Mbh\over3\times10^6\Mo}\right)^{1/4}.
\end{eqnarray}
A time-scale of paramount importance for the purpose of this paper is
the two-body relaxation time (see Eq. \ref{e:tr}). In this paper, time-scales will be compared to a related ``reference time\footnote{We follow the definition of \citet{Sha78}. This reference time has a slightly different normalization than the relaxation time. A \citet{Bah76} cusp forms on a time-scale considerably shorter than $T_0$, see also equation (\ref{e:DyyBW}).}'',

\begin{eqnarray}\label{e:tau0}
T_{0} &\equiv& {3\pi\over 16}{v_0^3\over (Gm)^2n_0 \ln\Lambda}\nonumber\\
&=&22\Gyr\left({\Mbh\over3\times10^6\Mo}\right)^{5/4},
\end{eqnarray}
where the dependence of the Coulomb logarithm on $\Mbh$ was neglected
in the last line.

In the vicinity of a MBH ($r<r_{h}$), orbits are assumed to be
Keplerian, and the number of stars $N(E)dE$ in the energy range $(E,E+dE)$
is related to the DF by

\begin{equation}\label{e:Nf}
N(E)=\pi^{3}\sqrt{2}(G\Mbh)^{3}E^{-5/2}f(E).
\end{equation}

\citet{Bah76} showed that the DF for bound stars in steady state is \citep[see also][]{Sar06}

\begin{equation}\label{e:NfBW76}
f(E)\propto E^{1/4}\quad \to\quad N(E)\propto E^{-9/4}.
\end{equation}

If $N_{0}$ is the total number of stars that are bound to the MBH with
energies larger than $v_0^2$, then number of stars in the
interval $(\ln E, \ln E+d\ln E)$ is given by

\begin{equation}\label{e:dNdlnE76}
{dN\over d \ln E} = {5\over 4}N_{0}\left({E\over v_{0}^2}\right)^{-5/4}.
\end{equation}

The density of stars is related to the DF by
\begin{eqnarray}
n(r)&\!=\!&4\sqrt{2}\pi\int_{-\infty}^{G\Mbh/r}\!dE f(E)\sqrt{{\frac{G\Mbh}{r}}-E}\nonumber\\
&\propto&r^{-7/4}\quad(f\propto E^{1/4})\,\label{e:nr}
\end{eqnarray}

The critical scale for the semi-major axis of binaries at a distance
$r_0$ of the MBH is set by the hard/soft boundary in equation
(\ref{e:hs}). Setting $\xi=1$ and estimating the squared velocity
dispersion to be $v_{\rm disp}^2=G\Mbh/r_0$, this gives

\begin{equation}\label{e:ah}
a_{h} = 0.07\AU\left({\Mbh\over3\times10^6\Mo}\right)^{-1/2}.
\end{equation}
Larger MBHs live in systems with larger velocity dispersions, and
hence require tighter binaries in order for them to survive in their
vicinity.

We end this section by summarizing some of the assumptions made throughout the paper. The assumptions are further discussed at several places in the text.

All stars are single mass, Newtonian point particles, orbiting a static MBH that is much more massive than the stars. Binary stars interact with single stars, the latter having a fixed distribution as in equation (\ref{e:NfBW76}). Binary-binary interactions are neglected, as is dynamical friction. We consider only the evolution of the binaries that are bound to the MBH and assume a reservoir of unbound binaries that do not evolve. The properties of the reservoir are related to those of the MBH through the relation between mass and velocity dispersion (\ref{e:Msigma}). Evolution of energy and angular momentum is entirely due to uncorrelated, weak two-body interactions, such that recoil after an encounter of a binary with a single star is neglected, as are secular effects such as resonant relaxation. The internal semi-major axis of the binaries evolves only as a result of interactions with single stars, and we use cross-sections found in three-body calculations performed by other authors.

\section{Evolution of energy and angular momentum of the binaries}\label{ss:EJ}

In this subsection we consider the evolution of the energy $E$ and angular momentum $J$
of the center of mass of the binaries. If $N(E, J)dEdJ$ is the number
of stars in a volume of size $dEdJ$ around a point $(E, J)$, the
Fokker-Planck equation can be written as

\begin{eqnarray}\label{e:fpEJ}
{\partial N\over \partial t} &=& - {\partial \over\partial  E}\left[D_{E}N\right] + {1\over2}  {\partial^2 \over\partial E^2}\left[D_{EE}N\right]\nonumber\\
&& -  {\partial \over\partial J}\left[D_{J}N\right] + {1\over2}  {\partial^2 \over\partial J^2}\left[D_{JJ}N\right]\nonumber\\
&& + {\partial^2 \over\partial E\partial J}\left[D_{EJ}N\right]
\end{eqnarray}

The notation can be simplified by introducing dimensionless quantities

\begin{equation}\label{e:dimless}
\tau = t/T_{0};\quad y = E/v_0^2;\quad j = J/J_c(E),
\end{equation}
where $J_{c}(E) =  G\Mbh(2 E)^{-1/2}$ is the angular momentum of a
circular orbit of energy $E$. Equation (\ref{e:fpEJ}) then becomes

\begin{eqnarray}\label{e:fpyj}
{\partial N\over \partial \tau} &=& - {\partial \over\partial  y}\left[D_{y}N\right] + {1\over2}  {\partial^2 \over\partial y^2}\left[D_{yy}N\right]\nonumber\\
&& -  {\partial \over\partial j}\left[D_{j}N\right] + {1\over2}  {\partial^2 \over\partial j^2}\left[D_{jj}N\right]\nonumber\\
&& + {\partial^2 \over\partial y\partial j}\left[D_{yj}N\right]
\end{eqnarray}

This partial differential equation is in general non-linear: the diffusion
coefficients depend on the stellar distribution, and this distribution
is in turn affected by the diffusion coefficients.

Direct numerical integration of the non-linear Fokker-Planck equations
is possible, but not straightforward \citep[see][]{Coh78}. The system
considered here is more complicated than the Fokker-Planck equations here describe,
since evolution of the internal parameters of the binary (semi-major
axis, exchange probability) is also followed. Instead of solving the
full problem, we assume simple forms for the diffusion coefficients which are determined by a fixed population of single stars
that has a \citet{Bah76} cusp (see \S\ref{s:scaling}). We then follow
the evolution of binary stars as they interact with the single
stars. This leads to a considerable simplification of the equations,
which now become linear.

Another simplification we will make is that three-body effects on $(E, J)$ evolution are neglected. After a strong encounter, a hard binary recoils with a velocity $\sim \sqrt{G m/a}$, which can lead to its ejection. However, for every such ejection, there are of order $\sim \ln \Lambda$ ejections at similar velocities due to the cumulative effect of weak encounters\footnote{We only follow the bound binaries. For each binary ejected with velocities much larger than the escape velocity, several binaries will be ejected at lower velocities, so the effect of strong encounters on the binary population is in fact even smaller than suggested.}. The energy given in strong encounters to single stars is also smaller by order $1/\ln\Lambda$ than that due to weak interactions, so our assumption of a static background of is not affected by the negligence of strong encounters. This is analogous to the role of strong single-single encounters on the dynamics, which was shown to be unimportant \citet{Fre06}. The assumption is further justified by the fact that very little hardening of binaries is observed in our simulations (see figure \ref{f:a}).

\subsection{Diffusion in energy}\label{s:DEE}

The random motion in energy space due to two-body interactions leads to energy diffusion.
We make the simple assumption that the single
stars are distributed according to the solution found by
\citet{Bah76}, $f_{s}(y)\propto y^{1/4}$. The energy diffusion term $D_{yy}$ was estimated analytically by
\citet{Lig77} to be 

\begin{equation}\label{e:DyyBW}
D_{yy} = 85 y^{9/4}
\end{equation}
(see their equation 51b for $p=1/4$). Intuitively, the power $9/4$ can
be estimated by noting that the relaxation time is inversely
proportional to the DF, $t_r(y)\propto y^{-1/4}$, and thus $D_{yy}\sim y^2/t_r(y)\sim
y^2y^{1/4}\sim y^{9/4}$.

A detailed analysis shows that in steady state, the two-body drift term
vanishes \citep{Bah76, Lig77}. This result was obtained for single mass
stars. Since our binaries are twice as massive as the single stars,
there will be some drift in this case because dynamical
friction becomes more important. Nevertheless, we neglect this effect
for simplicity, setting $D_{y}  = 0$.

\subsection{Diffusion in angular momentum}\label{s:DJJ}

The angular momentum diffusion term $D_{jj}$ was estimated analytically by
\citet{Lig77} to be 

\begin{equation}\label{e:DjjBW}
D_{jj} = 6.6 y^{1/4}
\end{equation}
(see their equations [55] and [56] for $p=1/4$). Intuitively, the
relaxation time is the time-scale for the angular momentum to change
by order of the circular angular momentum, or for $j$ to change by
order unity.

The angular momentum drift term due to two-body interactions is related
to the angular momentum diffusion term by

\begin{equation}\label{e:Dj2}
D_{j} = {D_{jj}\over 2j}
\end{equation}
\citep{Lig77}.

\section{Internal evolution of the binaries}\label{s:bin}

As the orbit of the center of mass of the binary with respect to the
MBH changes due to encounters with other stars, the characteristics of
the binary itself also change as a result of such interactions. In the model used in
this paper, the most important quantity that evolves due to
interactions with stars is the semi-major axis of the binaries. For
hard binaries ($\xi\gg1$, see equation \ref{e:hs}), the binary
typically shrinks in interactions with single stars. The binary thus
becomes even harder, and can potentially survive at closer distances
with respect to the MBH. Soft binaries ($\xi\ll1$) become softer, and
will eventually be ionized. In addition to the change in the
semi-major axis, one of the components of a binary may swap with the
incoming single star during the interaction. The probability that an
exchange interaction has occurred is followed during the evolution.

\subsection{Evolution of the semi-major axes of the binaries}\label{ss:a}

\citet{Heg93} find the probability
$P(\Delta|v)$ that a binary with initial energy $\epsilon>0$ that
interacts with a star with dimensionless velocity

\begin{equation} \label{e:v}
v = {v_3\over v_c}
\end{equation}
has a relative change in velocity

\begin{equation}\label{e:del}
\Delta\equiv {\epsilon' - \epsilon\over \epsilon},
\end{equation}
where $\epsilon'$ is the new energy of the binary (hardening implies
$\Delta>0$, softening implies $\Delta<0$; the convention used here is
that bound binaries have positive energy $\epsilon$). In equation
(\ref{e:v}), $v_3$ is the velocity of the incoming star in the center
of mass of the system, and the critical velocity $v_c$ is the minimal
velocity required to disrupt the binary, i.e.,

\begin{equation}\label{e:vc}
\epsilon = {1\over3}m v_c^2;\quad v_c = \sqrt{3\epsilon\over m}
\end{equation}
\cite[see][just before equation 2]{Heg93}.

The dimensionless probability $P(\Delta|v)$ is related to the
differential cross-section for relative energy changes $\Delta$ by

\begin{equation}
P(\Delta|v) = {v^2\over \pi a^2}{d\sigma\over d\Delta}.
\end{equation}
This probability function was evaluated through numerical three-body
integrations by \citet{Heg93}, and a fit to the results of those
simulation is given in equations (10, 35, 36, 37, 38) of that
paper.

The rate at which binaries are scattered from energy $\epsilon$ to
$\epsilon'$ depends on the DF of the densities and
velocities of the stars around it. For the model of a galactic nucleus
assumed here, the DF is given by $f(E)=A
E^{1/4}=A[G\Mbh/r-v_3^2/2]^{1/4}$ (see equation \ref{e:NfBW76}). The
DF is normalized such that $\int d^3 v f(E)=n(r)$, where $n(r)$ is the
number density of (single) stars. The rate at which the energy of a
binary changes from $\epsilon$ to $\epsilon'$ is then

\begin{equation}
Q(\epsilon, \epsilon') = {4\pi A\over \epsilon} \int^{\sqrt{2G\Mbh/r}} v_3^2 dv_3 \left[{G\Mbh\over r} - {1\over2}v_3^2\right]^{1/4} v_3 {d\sigma\over d\Delta}.
\end{equation}
The upper limit of the integral is dictated by $f=(E<0)=0$; for the lower limit, see below. Using Eq. (\ref{e:v}) to make the velocities dimensionless, and
eliminating $d\sigma/d\Delta$ in favor of $P(\Delta|v)$ gives

\begin{eqnarray}\label{e:Qdimless}
Q(\epsilon, \epsilon') &=& 4\pi^2 A a^2 v_c^{9/2}{1\over\epsilon}\\
&&\times \int_{v_{\rm min}}^{\sqrt{2G\Mbh/(rv_c^2)}} dv v \left[{G\Mbh\over rv_c^2} - {1\over2}v^2\right]^{1/4}P(\Delta|v).\nonumber
\end{eqnarray}

This expression can be further simplified by using 
equation (\ref{e:vc}), and by defining

\begin{equation}\label{e:x}
x\equiv {\epsilon r\over G\Mbh m}.
\end{equation}
The expression for $Q$ which is then obtained is

\begin{equation}\label{e:QA}
Q(\epsilon, \epsilon') =  3^{9/4}\pi^2AG^2m^{7/4}\epsilon^{-3/4} R(x, x'),
\end{equation}
where
\begin{equation}\label{e:Rxx}
R(x, x') = \int_{v_{\rm min}}^{\sqrt{2/3x}}dv v \left[{1\over3x} - {1\over 2}
v^2\right]^{1/4}P(\Delta|v).
\end{equation}
If the binary hardens, the lower limit of the integral is zero. If the
binary softens ($\Delta<0$), the maximal amount of energy that the
binary can gain is $v^2$, because the intruder has not more kinetic
energy. As a result, the lower limit of the integral is given by
\begin{equation}\label{e:vmin}
v_{\rm min} = \left[\max(0, -\Delta)\right]^{1/2}.
\end{equation}
The upper limit of the integral follows from the assumption that the
single stars are bound to the MBH.

To find $A$, we normalize such that at $r=r_h=G\Mbh/v_{0}^2$

\begin{eqnarray}\label{e:nh}
n_{0} &=& 4\pi A\int_0^{\sqrt{2G\Mbh/r}} v^2 dv \left[{G\Mbh\over r_h} - {v^2\over2}\right]^{1/4}=8.88  Av_{0}^{7/2}, \nonumber\\
\end{eqnarray}
so that
\begin{equation}\label{e:A}
A = {n_{0}\over8.88v_{0}^3}v_{0}^{-1/2}.
\end{equation}
Combining (\ref{e:QA}) and
(\ref{e:A}) then gives

\begin{equation}\label{e:QoverR}
Q(\epsilon, \epsilon') = 13.2{(Gm)^2n_{0}\over v_{0}^3}\left({\epsilon\over mv_{0}^2}\right)^{-3/4}{1\over mv_{0}^2} R(x, x').
\end{equation}

Rather than considering all the possible energy changes for a given
value of $x$, we make a Fokker-Planck approximation (see also appendix A) in
which only the first and the second moments of $Q$ are considered, so

\begin{eqnarray}\label{e:Dn}
\langle \Delta^n R\rangle &=& \int_{\Delta_{\rm min}}^{\infty}d\Delta \Delta^n R \\ 
&=& \int_{\Delta_{\rm min}}^{\infty}d\Delta \Delta^n \int_{v_{\rm min}}^{\sqrt{2/3x}}dv v \left[{1\over3x} - {1\over 2}
v^2\right]^{1/4}P(\Delta|v)\nonumber
\end{eqnarray}
The most negative relative energy change for the binary that is
considered is $\Delta=-1$, because the binary becomes unbound for
larger changes. Also, the most negative energy must be smaller than
the energy of the stars moving at the escape velocity at a given
distance from the MBH. This leads to the lower limit

\begin{equation}\label{e:Delmin}
\Delta_{\rm min} = \max\left[-1, -\left({2\over3x}\right)^2\right]
\end{equation}
of the integral in equation (\ref{e:Dn}). Numerical integration and
fitting then yields the following results for $0.03<x<300$:

\begin{eqnarray}\label{e:fitDEL2}
&&\ln(\langle \Delta^2 R\rangle) =  \sum_{i=0}^{3} a_i\left(\ln x\right)^i;\nonumber\\
&&(a_0, a_1, a_2, a_3) = ( -1.03, -1.23, 0.0456, -0.00729)\nonumber\\
\end{eqnarray}

\begin{eqnarray}\label{e:fitDELpos}
&&\ln(\langle \Delta R\rangle) = \sum_{i=0}^{3} b_i\left(\ln x\right)^i\quad (\langle \Delta R\rangle > 0); \nonumber\\
&&b_0 = -2.01;\quad b_1 = -0.0113;\quad b_2 = -0.353;\quad b_3 = 0.0312.\nonumber\\
\end{eqnarray}

\begin{eqnarray}\label{e:fitDELneg}
&&\ln(-\langle \Delta R\rangle) = \sum_{i=0}^{3} c_i\left(\ln x\right)^i\quad (\langle \Delta R\rangle < 0); \nonumber\\
&&c_0 = -5.488;\quad c_1 = -6.99; \quad c_2 = -2.44; \quad c_3 = -0.356.\nonumber\\
\end{eqnarray}

\begin{figure}[!h]
\includegraphics[angle=270,scale=.37]{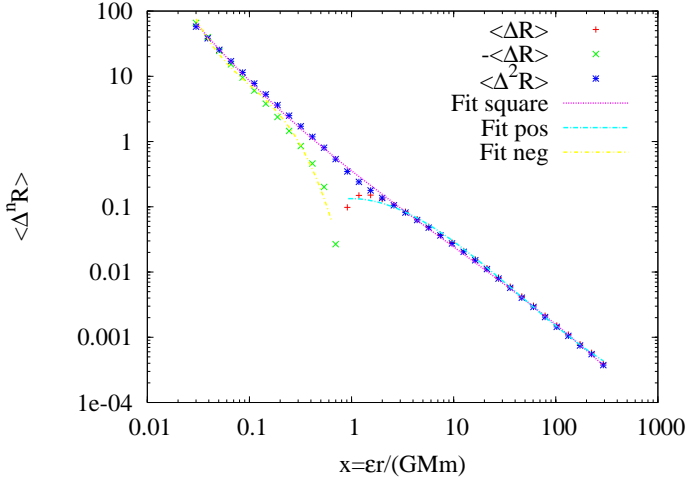}
\figcaption{The Fokker-Planck coefficients in equation (\ref{e:Dn})
and the fits (\ref{e:fitDEL2} - \ref{e:fitDELneg}).\label{f:DELn}}
\end{figure}

The results from the integration and the fits are displayed in figure
(\ref{f:DELn}).

\subsection{Exchange interactions}\label{ss:ex}

After an interaction between a binary and a single star that does not
disrupt the binary, the new binary may be composed out of the same
stars it started with before the interaction, or one of the the
original components may be exchanged for the single star. This is of
importance, since it provides a mechanism that is likely to
exchange main sequence stars for compact remnants, which may lead to
the formation of X-ray binaries. The relevance of this mechanism is
highlighted by globular clusters, in which the rate of interactions of
hard binaries correlates tightly with the number of X-ray sources
\citep{Poo03, Fre08} and cataclysmic variables \citep{Poo06}.

The cross-section for an exchange interaction to happen for a soft
binary is found to be \citep[][see equation 5.3]{Hut83}

\begin{equation}\label{e:sigsex}
\sigma_{s}^{\rm ex} = {160\over81}\pi (Gm)^2m^2\epsilon^{-2}v^{-6};\quad (v\gg1).
\end{equation}
The dimensionless velocity $v$ was defined in equation (\ref{e:v}). The
exchange cross-section for a hard binary may be estimated by
\citep{Heg96}

\begin{equation}\label{e:sighex}
\sigma_{h}^{\rm ex} = \bar{\sigma}{\pi\over4}(Gm)^2m^2\epsilon^{-2}v^{-2};\quad\bar{\sigma}=4.5\quad (v\ll1).
\end{equation}
In these equations $v_3$ is the relative velocity, and $v$ is defined
as in equation (\ref{e:v}). The numerical factor $\bar{\sigma}$ was
found by summation of the average of the single mass cross-sections
for direct and resonant exchange in table (1) from \citet{Heg96}.

These cross-sections can be interpolated to find the general
cross-section for any velocity,

\begin{equation}\label{e:sigex}
\sigma^{\rm ex} = \left[{1\over \sigma_{s}^{\rm ex}} + {1\over \sigma_{h}^{\rm ex}} \right]^{-1}.
\end{equation}

The rate of exchange interactions per binary is then

\begin{eqnarray}\label{e:exrate}
\Gamma_{\rm ex} &=& 4\pi\int v_3^2 dv_3 f_{s} v_{3}\sigma^{\rm ex}\nonumber\\
&=&53{(Gm)^2n_{0}\over v_{0}^3}\left({\epsilon\over m v_{0}^2}\right)^{1/4} R_{\rm ex}(x),
\end{eqnarray}

or in units of $T_0$,

\begin{equation}\label{e:exrateT0}
\Gamma_{\rm ex}T_{0} = 3 \left({10\over \ln\Lambda}\right)\left({\epsilon\over m v_{0}^2}\right)^{1/4}R_{\rm ex}(x).
\end{equation}

\begin{figure}[!h]
\includegraphics[angle=270,scale=.37]{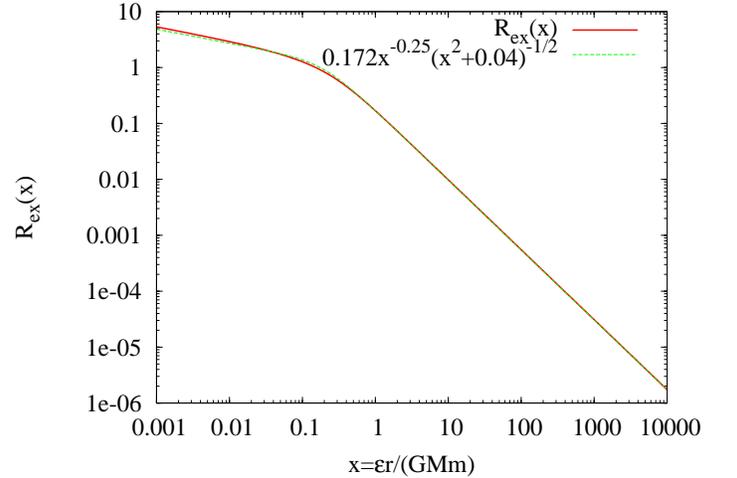}
\figcaption{The function $R_{\rm ex}$ in equation (\ref{e:Rex}) as a function of dimensionless binary energy $x= \epsilon r/(G\Mbh m)$. \label{f:Rex}}
\end{figure}

In these expressions,

\begin{equation}\label{e:Rex}
R_{\rm ex} \equiv \int_{0}^{\sqrt{2/3x}}dv v^3\left({1\over 3x} - {1\over2}v^2\right)^{1/4}\left({4\over\bar{\sigma}}v^2 + {81\over160} v^6\right)^{-1}.
\end{equation}
For this derivation similar arguments and definitions were used as in
\S\ref{ss:a}. A fit to numerical integration of equation (\ref{e:Rex})
yields

\begin{equation}\label{e:Rexfit}
R_{\rm ex} = 0.172 x^{-1/4}(x^2+0.04)^{-1/2}.
\end{equation}
The function (\ref{e:Rex}) and the fit (\ref{e:Rexfit}) are plotted in
figure (\ref{f:Rex}).

In the simulation, the probability that a given binary has experienced
an exchange interaction is estimated as follows. During every
time-step $dt_{i}$ at a time $t_{i}$, the probability that there was no
exchange interaction is

\begin{equation}\label{e:Pnoex}
P_{i}^{\rm no\,\, ex} = 1 - \Gamma_{\rm ex}dt_{i}.
\end{equation}

The probability that a binary has experienced an exchange interaction
after a time $t=\Sigma_{i}dt_{i}$ is then

\begin{equation}\label{e:Pex}
P_{i}^{\rm ex} = 1 - \Pi_{i}P_{i}^{\rm no\,\, ex}.
\end{equation}

\section{Destruction of binaries}\label{s:destruct}
There are two major ways in which binaries can be destroyed. One
channel for binary destruction is that (soft) binaries interact with
field stars, and become gradually softer. If the binding energy is
less than some critical value, the two components in the binary are no
longer considered to be bound to each other. We will refer to this
process as ``ionization'', following common nomenclature
\cite[e.g.][]{Heg03}.

The other mechanism that destructs binaries is when a binary star
comes at peri-apse so close to the MBH that the tidal force exerted by
the MBH exceeds the internal force of the binary components. This happens when the peri-apse of the binary is smaller than the tidal radius,

\begin{equation}\label{e:rt}
r_{p}<r_{t}=\left({\Mbh\over 2m}\right)^{1/3}a.
\end{equation} 
This happens typically on highly eccentric orbits, so that the angular
momentum of the binary with respect to the MBH is smaller than the
angular momentum of the loss-cone,

\begin{equation}\label{e:Jlc}
J < J_{lc}=\sqrt{(1+e)G\Mbh r_{t}}.
\end{equation} 

Theoretically, a binary can also be disrupted when its energy with
respect to the MBH becomes larger than the energy associated with a
circular orbit with radius $r=r_{t}$, but this is extremely unlikely,
since the time-scale to reach $J<J_{lc}$ is much shorter \citep{Fra76,
Lig77}.

\subsection{The loss cone}\label{ss:lc}

When a binary has
$J<J_{lc}$, this does not necessarily imply that it will be disrupted,
since the angular momentum can be scattered while the star is going to
peri-apse. The event rate of loss-cone disruptions was first studied
by \citep{Fra76, Lig77}, and many later papers \citep[e.g.,][]{Bah77,
Mag99, Sye99}. A distinction is made between a ``full loss-cone''
regime, where the orbital angular momentum change is much larger than
the size of the loss-cone, and an ``empty loss-cone'' regime, where
this change is much less than the size of the loss-cone.

The change in angular momentum per orbit is $\Delta J_p\approx
\sqrt{P/D_{JJ}}$. If $\Delta J_p\gg J_{lc}$, the binary can jump in
and out of the loss-cone on a dynamical time, and the loss-cone in
this regime is always full. The dimensionless rate (in units of $T_0$)
at which a binary with semi-major axis $a$ and energy $E$ is disrupted
is then approximately

\begin{equation}\label{e:gammafull}
\gamma_{\rm full}=\left({J_{lc}\over J_c}\right)^2{T_0\over P}.
\end{equation}

The demarcation between the empty and full loss-cone regime can be
found by setting $\Delta J_p=J_{lc}$. Evaluating this at the radius of
influence gives (see equations \ref{e:tau0}, \ref{e:P0},
\ref{e:rhMsig}, \ref{e:Jlc})

\begin{equation}\label{e:aemptyfull}
a_{\rm full} = 0.1\AU  \left({\Mbh\over3\times10^6\Mo}\right)^{-5/6}.
\end{equation} 
If $a>a_{\rm full}$, then the binary is in the empty loss-cone regime throughout the entire cusp; the loss-cone becomes full only for stars at radii $>r_h$. In contrast, if $a<a_{\rm full}$, it is in the empty loss-cone regime close to the MBH, and in the full loss-cone regime in regions further away from the cusp.

In the empty loss-cone regime, any binary with $J<J_{lc}$ is
immediately disrupted. The dimensionless rate at which a given binary
enters the loss-cone is of order $1/t_r$, or, more precisely
\citep{Lig77}

\begin{eqnarray}\label{e:Gammaempty}
\gamma_{\rm empty} &=& {T_0\over t_r(E) \ln(J_{c}/J_{lc})}\nonumber\\
&=& \left({E\over v_0^2}\right)^{1/4}{1\over \ln(J_{c}/J_{lc})}.
\end{eqnarray}

Since the loss-cone introduces an scale to the system, it must be treated carefully in the MC simulations. See \S\ref{s:MC} and \citet{Sha78} for details.

\subsection{Ionization}\label{ss:ion}

The rate at which soft binaries lose energy due to the cumulative
effect of many weak encounters can be estimated as follows. The change
$\delta v$ of one of the components of the binary in an encounter with
a field star at a distance $b$ is $\delta v\sim Gm/bv_{\rm disp}$. On
average these encounters cancel, $\langle \delta v\rangle=0$, but the
root mean square of the velocity increases, so that $\langle \delta
\epsilon\rangle\sim\delta v^2$. The rate at which single stars pass
one of the binary components at a distance between $(b, b+db)$ is
$n_{s}v_{\rm disp}bdb$, so that the rate at which the energy of the
binary increases is $\dot{\epsilon}\sim (Gm)^2n_s\ln\Lambda/v_{\rm
disp}$. The ionization time for soft binaries $t_{\rm
ion}\sim\epsilon/\dot{\epsilon}$ is then

\begin{equation}\label{e:tion}
t_{\rm ion} \sim {v_{\rm disp}\over G n_{s}ma\ln\Lambda}.
\end{equation}
In the context of \S\ref{s:bin} the dimensionless ionization rate may
be estimated by the rate at which binaries change their semi-major
axis (harden/soften) in three-body interactions at a rate
(equations \ref{e:QoverR}, \ref{e:Dn})

\begin{equation}\label{e:gamma_a}
\gamma_{a}=0.3\left({10\over\ln\Lambda}\right)\left({\epsilon\over mv_{0}^2}\right)^{1/4} {\langle  \Delta^2R(x)\rangle \over\langle \Delta^2R(1)\rangle}.
\end{equation}
Here $\langle  \Delta^2R(1)\rangle\approx0.36$, and $\langle  \Delta^2R(x)\rangle$ is
approximately proportional to $(E/\epsilon)^{5/4}$. The rate
$\gamma_{a}$ can be interpreted as the ionization rate if $x\ll1$, or
as the hardening rate when $x\gg1$.

\section{Dynamical rates}\label{s:times}

The dynamical processes described in the previous sections can be
characterized by their rates, which provide insight in the results of
the simulations presented below. Here the dynamical rates are
summarized. All rates are given in units of $[1/T_{0}]$. Once the rates are known, a simple model can be made that describes the DF and the binary disruption rates. This model is in qualitative agreement with the results from the MC simulations and gives some insight in the dynamical processes.

\subsection{Rates}\label{ss:rates}

The rate for an energy change of order unity due to two-body
interactions is (Eq. \ref{e:DyyBW})

\begin{equation}\label{e:gamma_diff}
\gamma_{\rm diff} \equiv {D_{EE}T_{0}\over E^2} = 85\left({E\over v_{0}^{2}}\right)^{1/4},
\end{equation}
only weakly dependent on energy.

While hardening or softening, binaries have a probability to have an
exchange interaction during every three-body encounter. The characteristic
rate for these is
\begin{equation}\label{e:gx}
\gamma_{\rm ex} = 0.5 \left({10\over \ln\Lambda}\right)\left({\epsilon\over m v_{0}^2}\right)^{1/4}{R_{\rm ex}(x)\over R_{\rm ex}(1)} .
\end{equation}
In this expression, $R_{\rm ex}(1)\approx0.17$, and $\gamma_{\rm ex}\sim\,{\rm
const}$ for very soft binaries, and $\gamma_{\rm ex}\sim x^{5/4}$
for very hard binaries.

The hardening/softening rates was given in equation (\ref{e:gamma_a}), and the effective loss-cone rate is (see equations [\ref{e:gammafull}] and [\ref{e:Gammaempty}])

\begin{equation}\label{e:Gammalc}
\gamma_{\rm lc} = \min\left[\gamma_{\rm full}(E), \gamma_{\rm empty}(E)\right]
\end{equation}

Let us consider what is likely to happen to a star that enters the
cusp on a marginally bound orbit, $E\approx v_0^2$. If the binary is
very soft $(\epsilon\ll mv_{0}^{2})$, then the ionization rate
$\gamma_{\rm ion}$ is much larger than any of the other rates, and the
binary will quickly be ionized due to interactions with the other
stars. If the binary is hard, it is more likely to be tidally disrupted than to be
ionized. However, the tidal disruption rate is smaller than
the energy diffusion rate, and
it is more probable that the binary diffuses to higher energies with
respect to the MBH. While this is happening, the binary can harden
somewhat, but it is important to stress that the hardening rate is
always smaller than the diffusion rate. {\it This implies that when a
binary moves to higher energies, it becomes usually softer:} the
semi-major axis of the binary may decrease somewhat, but the velocity
dispersion of the stars in its vicinity grows more rapidly than the
circular velocity of the binary increases.

As the binaries diffuse to higher energies, some are tidally
disrupted. However, since in the empty loss-cone regime the tidal
disruption rate is always smaller than the energy diffusion rate,
$\gamma_{\rm empty}\lesssim \gamma_{\rm diff}$, this does not lead to
a very significant depletion of binary stars compared to the situation
that tidal disruptions are neglected; with respect to this process,
binaries could exist at all energies\footnote{This is no longer true
when resonant relaxation becomes important, see \citet{Hop06}}. This
is not true for ionization of binaries: since hardening occurs at a
smaller rate than energy diffusion, binaries with any energy $a$ will
eventually reach an energy $E_{s}(a)$ with respect to the MBH where
they become soft, and for energies $E\gg E_{s}(a)$, $\gamma_{\rm
ion}\gg \gamma_{\rm diff}$. As a result of this there will be an
exponential depletion of binary stars compared to the situation where
ionization is neglected.

The rate of exchange interactions for a given binary is comparable to
the hardening rate. This implies that $\gamma_{\rm ex}<\gamma_{\rm
diff}$, and therefore most binaries move first to higher energies and
are then ionized before they have experienced an exchange
interaction. The probability for an exchange interaction is therefore
low. This general result is confirmed in the MC
simulations below.

\subsection{Parametrized model}\label{ss:param}

The insight obtained from the dynamical rates presented here can be
used to make a simple analytical model of the DF
and the disruption rates. We present a model with several added parameters to fit the
results of the MC model as well as possible. The model does not accurately treat the full three dimensional dynamics, but does capture a number of important features, and helps to understand the underlying dynamics.

As argued before, loss-cone effects cannot affect the shape of the DF
by much for hard binaries, whereas an exponential cut-off is expected
for soft binaries. The DF can therefore be written as

\begin{equation}\label{e:fanalyt}
f(E) \propto E^{1/4}\exp\left[-E/(w_{s}\xi v_0^2)\right],
\end{equation}
where $w_s$ is a numerical parameter which we chose to fit the data
best. The normalized number of stars per $\log E$ is then

\begin{equation}\label{e:Nanalyt}
{1\over N_0}{dN\over d\ln E} = {5\over4} \left({E\over v_0^2}\right)^{-5/4}\exp\left[-E/(w_{s}\xi v_0^2)\right],
\end{equation}
where $N_0$ is the total number of binaries that are bound to the MBH (which is much smaller than the total number of single stars bound to the MBH). In these expressions, we use the initial hardness parameter $\xi$ the binary had when it entered the cusp. In reality, the internal energy of the binary evolves. However, since $\gamma_a \ll \gamma_{\rm diff}$ (equations \ref{e:gamma_a} and \ref{e:gamma_diff}), the internal energy of the binary in fact changes very little as it diffuses through the cusp.

The tidal disruption rate per $\log E$ as a function of energy is 

\begin{eqnarray}\label{e:tiddisrrate}
\Gamma_t &=& {5w_{t}\over4} \left({E\over v_0^2}\right)^{-5/4}\exp\left[-E/(w_{s}\xi v_0^2)\right]{\gamma_{lc}^2\over\gamma_{lc}+\gamma_{a}}
\end{eqnarray}
where $w_t$ is a parameter to better fit the results, and a factor $\gamma_{lc}/(\gamma_{lc}+\gamma_{a})$ accounts for the fact that the binary can also be disrupted by ionization, decreasing the probability that it will be tidally disrupted.

The ionization rates can be estimated as follows. For binaries with $\xi v_0^2 < E$, the rate is $\sim\gamma_a$. Binaries with $\xi v_0^2 > E$ are hard on circular orbits, but for a thermal distribution of eccentricities, a fraction of stars $\sim E/(w_{s}\xi v_0^2)$ spends a fraction of time $\sim \left [E/(w_{s}\xi v_0^2)\right]^{3/2}$ in regions where $G\Mbh /r>\xi v_0^2$, where $w_{s}$ is the same parameter as in equation (\ref{e:Nanalyt}). We therefore estimate the ionization rate to be

\begin{eqnarray}\label{e:iondisrrate}
\Gamma_i &=& {5w_{i}\over 4}  \left({E\over v_0^2}\right)^{-5/4}\exp\left[-E/(w_{s}\xi v_0^2)\right]\nonumber\\
&&\times\min\left[1, \left({E\over w_s \xi v_0^2}\right)^{5/2}\right]{\gamma_{a}^2\over\gamma_{lc}+\gamma_{a}}.
\end{eqnarray}
where $w_i$ is the third and last parameter to better fit the results.

Fitting by eye, we found that a reasonably good approximation for the numerical results is given by taking the parameters to be $(w_s, w_t, w_i) = (8, 1, 5)$ (see also figure \ref{f:andisr} below).

\section{Monte Carlo scheme}\label{s:MC}

The evolution in the previous sections been described in
the terms of Fokker-Planck equations. Such equations can be solved
using a MC method. The relation between diffusion equations and MC methods is reviewed in the appendix. An advantage of MC methods is that it is relatively straightforward to add effects such as the probability for exchange interactions
(\S\ref{ss:ex}) to such a method.

In a MC scheme, the DF is represented by a
number of binaries that represents a realization of the actual
distribution. This number of simulated binaries can be larger or
smaller than the actual number of binaries. At every time-step, all
monitored quantities for a given binary change, some deterministically
(if there is a drift), and some randomly (if there is diffusion). In addition, the
binary can be disrupted by loss-cone effects or three-body effects, there
is a probability that it experiences an exchange interaction, it can
become unbound, and so forth.

In the MC code, the energy and angular momentum of the binary are first updated according to

\begin{equation}\label{e:DEMC}
E(t + \delta t) = E(t) + \chi_{1} \left[ D_{EE} \delta t\right]^{1/2};
\end{equation}

\begin{equation}\label{e:DJMC}
J(t + \delta t) = J(t) + D_{J}\delta t + \chi_{2} \left[ D_{JJ} \delta t\right]^{1/2}.
\end{equation}
In these equations, $\chi_{1}$ and $\chi_{2}$ are standard normally distributed random variables that have a correlation $D_{EJ}/\sqrt{D_{EE}D_{JJ}}$. For details see \citet{Sha78}.

\begin{equation}\label{e:DepsMC}
\epsilon(t + \delta t) = \epsilon(t) + D_{\epsilon} \delta t + \psi \left[ D_{\epsilon\epsilon} \delta t\right]^{1/2},
\end{equation}
with $\psi$ an independent standard normal random variable, and

\begin{equation}
D_{\epsilon} = {7.8\over \ln\Lambda}\left({\epsilon\over m v_0^2}\right)^{1/4}\langle \Delta R\rangle;\quad D_{\epsilon\epsilon} = {7.8\over \ln\Lambda}\left({\epsilon\over m v_0^2}\right)^{1/4}\langle \Delta^2 R\rangle
\end{equation}
(see equations \ref{e:QoverR} --- \ref{e:fitDELneg}).

Once the change in internal energy is calculated, we update the probability that there has not yet been an exchange interaction for that binary using equation (\ref{e:Pnoex}).

A complication in MC simulations is that often the
probability distribution varies by orders of magnitude over the range
of interest. This implies that a very large number of particles is
required in order to resolve the probability distribution at values
where it is low, and hence that much more particles are present at
values where it is high. In the situation that is considered here, the
number of stars per unit energy, $dN/dE$, changes very steeply near a
MBH. For example, for a cusp with DF $f\sim E^{p}$,
$dN/dE\propto E^{-5/2+p}$. This implies that if one were to set up a
MC simulation naively, a very large number of particles needs
to be present at energies of order $E\sim1$, in order for there to be
only a few particles at large energies.

This problem can be resolved by using a ``cloning scheme'', as
introduced by \citet{Sha78}. We use a simplified version of their
scheme here, which we now briefly summarize.

All binaries start at small energies $E_{\rm new} = 0.5 v_{0}^2$, with
angular momentum drawn from an isothermal distribution in the range
$J/J_c=(0, 1)$, and with a semi-major axis drawn from a distribution
that depends on the model (see next section). For each individual
particle an adaptive time-step is computed, following the prescription
presented in \citet{Sha78}, and additional criteria to ensure that the
fractional changes in the binary properties are smaller than 0.01.

As the energy of the binary with respect to the MBH changes because of
two-body and three-body scattering, it occasionally diffuses to higher
energies. If the binary crosses an energy $E_{{\rm clone}, n} = 10^n$,
where $n=(0, 1, 2, 3)$, 9 identical binaries are made. All these
binaries are tagged to be clones. The binaries then proceed to evolve
independently. When a clone crosses the boundary $E_{{\rm clone}, n}$
at which it was produced, the binary stops to exist and is discarded
in the analysis. A binary can also be annihilated by physical
processes. The channels for annihilation are disruption by the
loss-cone; ionization; crossing a large, maximal energy $E_{\rm
max}$; and interactions that lead to an orbit that is unbound to the
MBH, where binaries are only considered to be ``bound'' by the MBH if
their energies are larger than $E_{b} = 0.4v_{0}^2$. We consider a binary to be ionized when $\epsilon/m_b E < 10^{-1}$. In case of these
physical annihilations, the process that destroys the binary is
recorded, so that it is possible to distinguish between ionizations
and loss-cone captures. If the destroyed particle is a clone, it is
not replaced. If it is one of the original particles, it is replaced
by a new particle with energy $E_{\rm new}$. In this way, the number
of particles in the simulations that are not clones remains constant
by construction.

Using this method, there are approximately equal numbers of simulated
binaries at all energies. There is no over-sampling of the low energy
region, and there are enough particles at high energy regions in order
to resolve those. At high energies, however, particles have less
statistical weight: any binary with energy $10^{n}<E/v_{0}^2 <10^{n+1}$ has a weight $w=10^{-n}$. This weight is
then taken into account when calculating the DF,
disruption rates, distribution of semi-major axes, etc. Typically,
1000 particles are used at the start of the simulation, which
procreate to a final number of about 5000 (depending on the
simulation). The distribution reaches steady state on a time-scale of the order of the relaxation time. These simulations typically completed in a few hours.

\section{Results}\label{s:results}

\begin{center}
\begin{table}[t]
\caption{Model parameters and GW rates}\label{t:model}
\begin{tabular}{llllll}
\hline
\hline
Name       &   $\Mbh $            &  $a_{\rm min}$     & $a_{\rm max}$     & $\xi_{0}$  & Comments\tabularnewline
           &   $[\Mo]$            &  [AU]              &  [AU]             &            &\tabularnewline
\hline
I          &  $3\times10^{6}$     &  0                 & 0                 &            &No $J$ and $a$ evolution \tabularnewline
IIa        &  $3\times10^{6}$     &  0.24              & 0.24              & 0.3        &\tabularnewline
IIb        &  $3\times10^{6}$     &  0.073             & 0.073             & 1          &\tabularnewline
IIc        &  $3\times10^{6}$     &  0.0073            & 0.0073            & 10         &\tabularnewline
IId        &  $3\times10^{6}$     &  $7.3\times10^{-4}$& $7.3\times10^{-4}$& 100        & \tabularnewline
IIIa       &  $1\times10^{3}$     &  13                & 13                & 0.3        &\tabularnewline
IIIb       &  $1\times10^{3}$     &  4.0               & 4.0               & 1          &\tabularnewline
IIIc       &  $1\times10^{3}$     &  0.4              & 0.4              & 10         &\tabularnewline
IIId       &  $1\times10^{3}$     &  0.04             & 0.04             & 100        & \tabularnewline
IV         &  $3\times10^{6}$     &  0.01              & 0.3               & 0.24-7.3   &Favored MBH model \tabularnewline
V          &  $1\times10^{3}$     &  0.01              & 10                & 0.4-400   &Favored IMBH model \tabularnewline
\hline
\multicolumn{6}{l}{Parameters for the simulations. Models I and II are given for theoretical purposes;}
\tabularnewline
\multicolumn{6}{l}{note that some of the semi-major axes in those models cannot hold for MS stars.}\tabularnewline
\label{t:t1}
\end{tabular}
\end{table}
\end{center}

A number of different models are considered to probe how the different
effects of binary evolution, loss-cone disruptions, and the dependence
on the initial conditions modify the result. The different models are
summarized in table (\ref{t:model}).

Model I considers only energy $E$ evolution. The semi-major axis and
angular momentum of the binaries does not evolve, and there is no
loss-cone. This reproduces the
\citet{Bah76} distribution to good approximation, and is used for comparison with other
models.

Models IIa --- IId have a $\Mbh=3\times10^{6}\Mo$ MBH, and include all
the dynamical effects discussed in the paper (such as $J$ and $a$
evolution, loss-cone). For each of these models,
the binaries have initially all the same semi-major axis. This makes
it possible to compare the dynamical evolution of different
binaries. The semi-major axes are chosen such that the hardness ratio
equals $\xi_{0}\equiv Gm/2av_{0}^2=(0.3, 1, 10, 100)$ at the radius of
influence (see equation \ref{e:hs}). We note that these values were chosen to study the dynamics, and not to be necessarily realistic. For example, some of these values imply that $a<\Ro$; these cases can be thought of as representing compact remnants, or as for theoretical interest.

The same assumptions are made for models IIIa --- IIId, with the only
difference being that the mass of the MBH was now chosen to be
$\Mbh=10^3\Mo$, as is perhaps the situation in some globular clusters.

Models IV and V are the models that are considered to be the most
realistic representation of the nucleus of a Galaxy and of globular
clusters, within the limits of our model. They again include all the
dynamical effects. The initial distribution of the binary semi-major
axes was chosen to have the simple form $\log a$, $dN/d\ln
a\propto\,\, {\rm const}$, which is roughly consistent with the log-normal period distribution of binaries in the Solar neighborhood \citep{Duq91}. For the Galactic nucleus model, the minimum
is $a_{\rm min}=0.01\AU$, and the maximum $a_{\rm max}=0.3\AU$, in
accordance with the range of semi-major axes used by
\citet{YuQ03}. The minimal semi-major axis is the distance for Solar
type stars to nearly touch each other. The maximal semi-major axis is
chosen such that $\xi_{0} \approx 0.3$; binaries wider than that will quickly dissolve. For the globular cluster
model, the minimal semi-major axis is the same, but the maximal
semi-major axis is $a_{\rm max}=10\AU$, so that $\xi_{0}\approx0.5$.

The figures of models I-V all show on the left hand side the case of
$\Mbh=3\times10^{6}\Mo$, and on the right hand side the case of
$\Mbh=10^{3}\Mo$. Top figures show models II and III, where all
binaries start with the same semi-major axis, while bottom figures
show the favored models.

\subsection{The distribution function}\label{ss:df}

\begin{figure*}[!htb]
    \begin{center}
         \begin{tabular}{c c}
              \scalebox{0.35}{\includegraphics[angle=270]{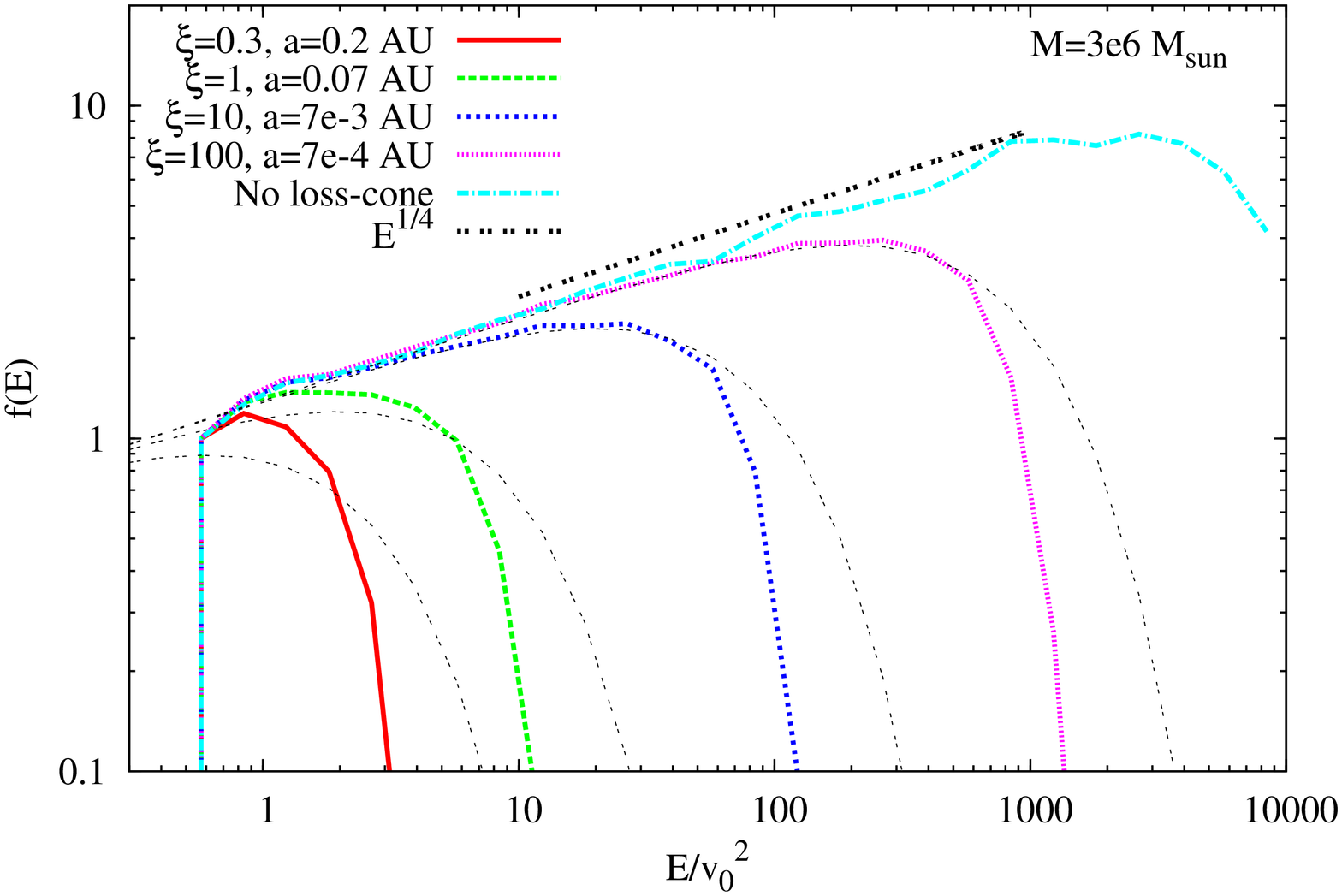}} &
              \scalebox{0.35}{\includegraphics[angle=270]{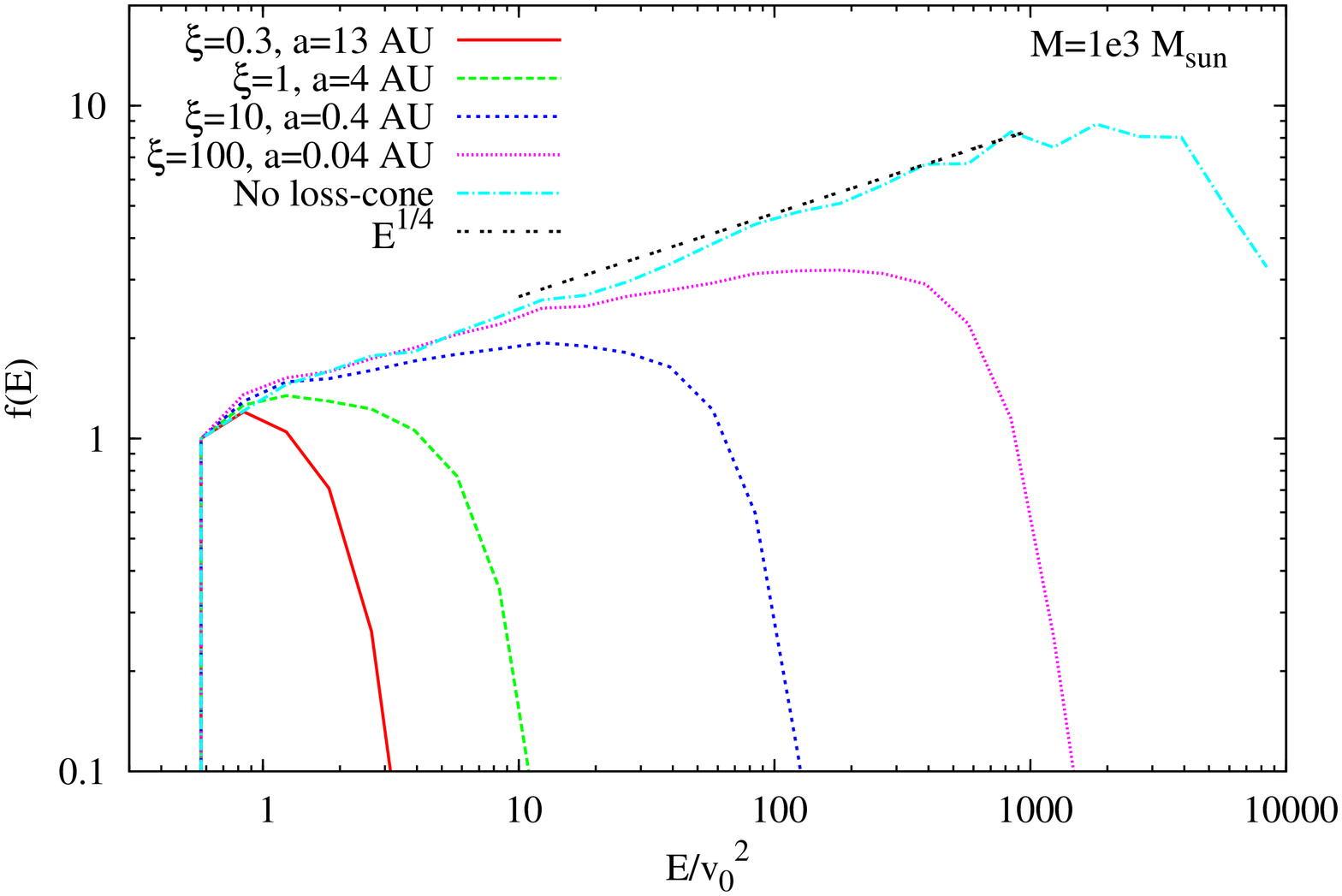}} \\
              \small \emph{(a)} & \small \emph{(b)}\\
              \scalebox{0.35}{\includegraphics[angle=270]{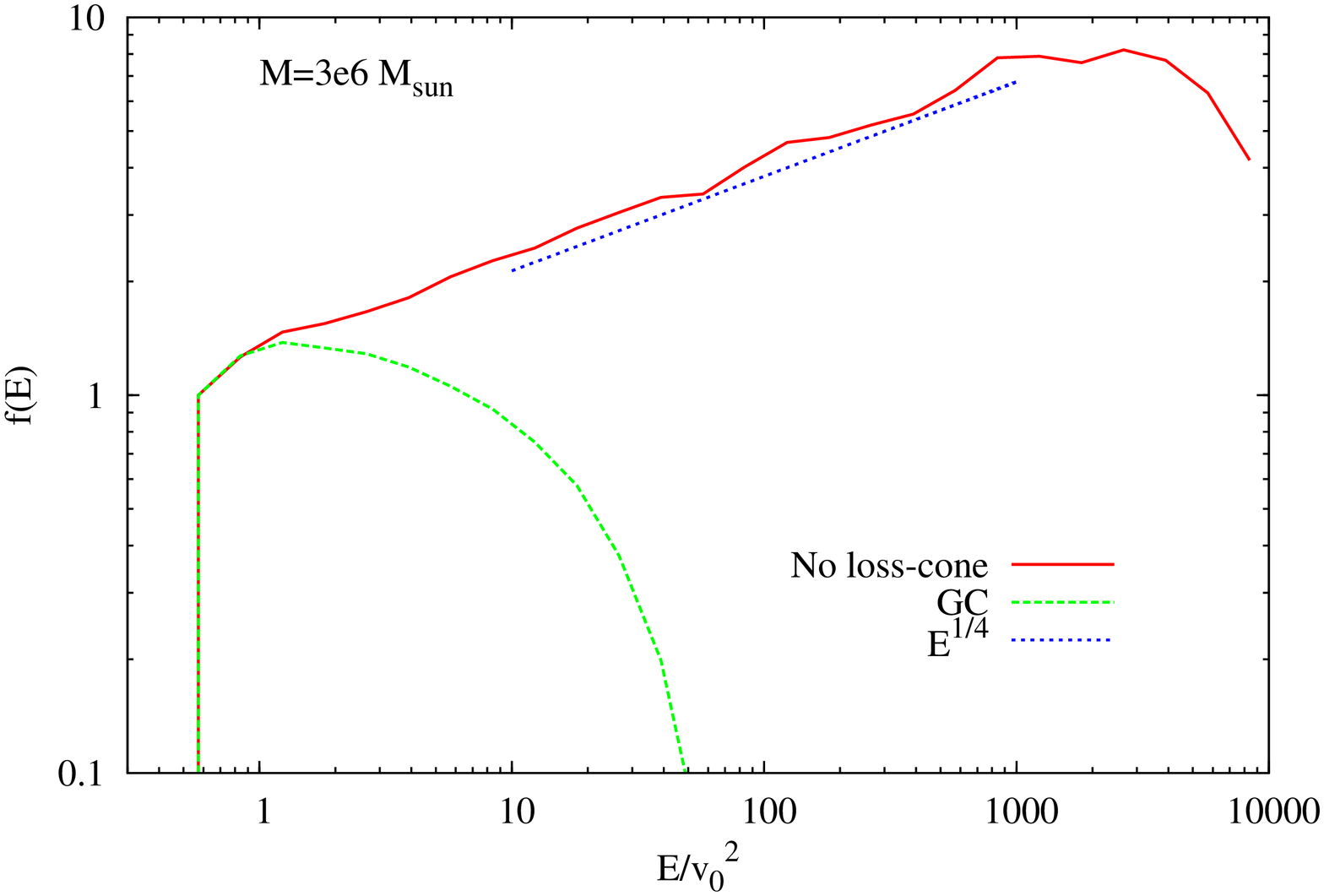}} &
              \scalebox{0.35}{\includegraphics[angle=270]{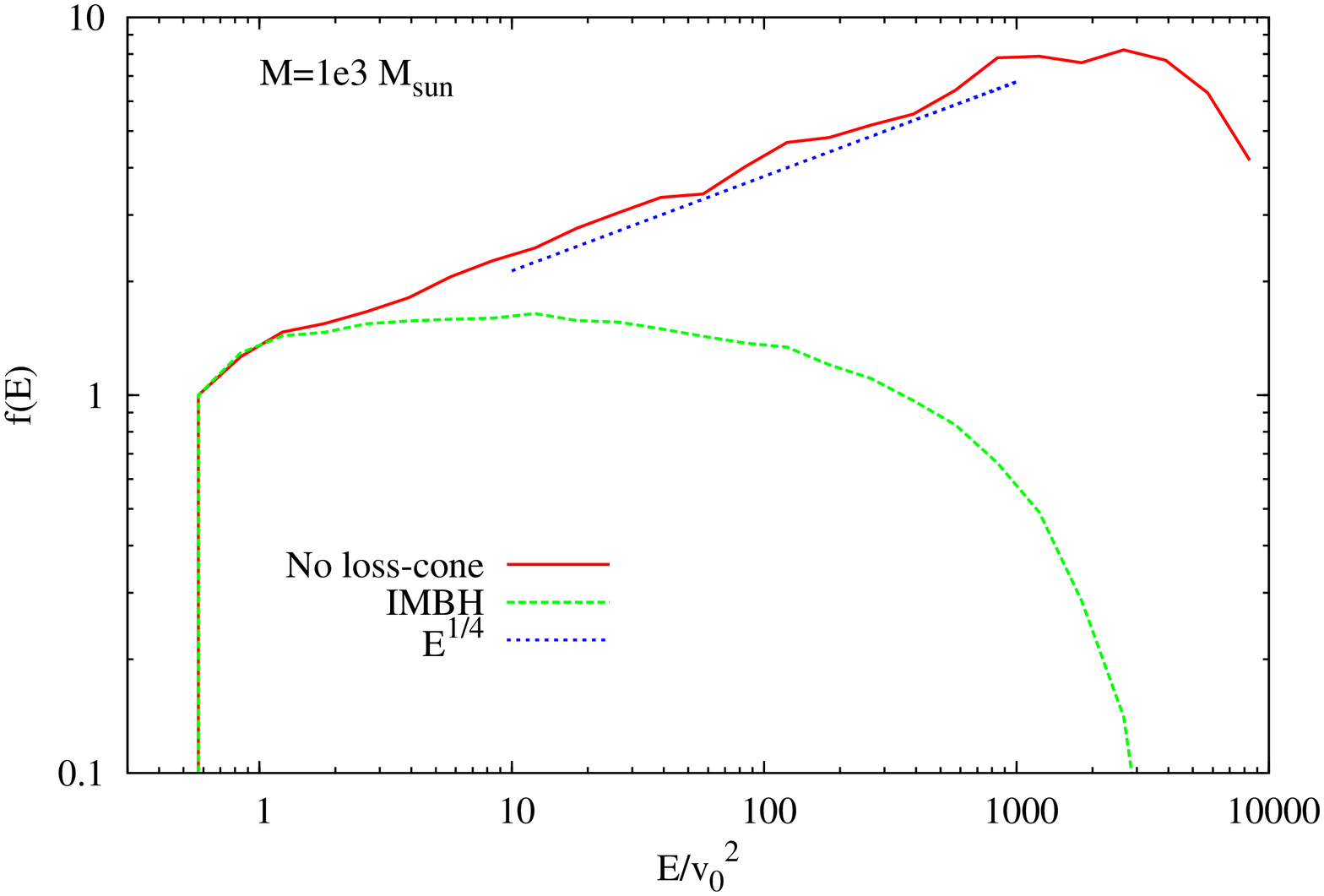}}\\
              \small \emph{(c)} & \small \emph{(d)}\\
         \end{tabular}
      \caption{DF of several types of binaries
      near MBHs and IMBHs, as a function of energy. Figure ({\it a})
      shows models II, figure ({\it b}) shows models III,
      figure ({\it c}) shows model IV, and figure ({\it c}) shows
      model V. In all figures we also show model I in which there are no tidal disruption of binary evolution effects for comparison. The DF is strongly affected by the environment near
      MBHs, which is hostile to binaries. Soft binaries are more
      easily disrupted than hard binaries; the latter can survive to
      somewhat larger energies. In figures ({\it a}) the analytical approximation given in equation (\ref{e:fanalyt}) is also plotted. The straight line is $\propto E^{1/4}$, which is the
      theoretical \citet{Bah76} result when there are no loss-cone and
      ionizations. \label{f:fE}}
    \end{center}
\end{figure*}

Figure (\ref{f:fE}) shows the DF $f(E)$, integrated
over all angular momenta for several different examples. If tidal disruptions and ionizations are
neglected for distances much larger than $r_t$, a $f(E)\propto E^{1/4}$ \citet{Bah76} profile is obtained,
which provides also a test for the MC code. This profile is modified
for cases where the loss-cone and three-body interactions are
included. Binaries become ionized when the reach high energies, and
there is an exponential cut-off to the DF. For the
cases where at the radius of influence $\xi=0.3$ (models IIa and
IIIa), the DF quickly drops to zero. It is clear that such
binaries can in fact not exist in the vicinity of MBHs. For smaller semi-major axes
($\xi\geq1$, models b-d), binaries can to some extend survive on
orbits bound to the MBH, and the smaller the semi-major axis, the
higher the energy to which it survives. This is the combined result of
the fact that tight binaries are less likely to enter the loss-cone of
the MBH, and they are less easily ionized. The qualitative comparison between the MC results and the parametrized analytical model, shown for $\Mbh=3\times10^6\Mo$, is good. Comparing the figures for $\Mbh=3\times10^6\Mo$ and $\Mbh=10^3\Mo$, it can be seen that the results are very similar. This is to be expected: as discussed in \S\ref{s:times}, the only process that can strongly modify the DF is the annihilation of binaries by ionization. The equations determining the rates for this to happen only contain the dimensionless quantity $\xi$ which relates the internal energy of the binary to the velocity dispersion around the MBH. In the figures we scale the energies with the velocity dispersion at the radius of influence, so that any dependence on the MBH mass is scaled out. The only MBH mass dependent process is then disruption by tidal effects: the critical distance where the loss-cone becomes full is $\Mbh$ dependent. However, in absence of resonant relaxation, as assumed here for simplicity, loss-cone effects cannot strongly modify the DF. The binary disruption rates will be further
discussed in figure (\ref{f:disr}). 

The bottom figures show the DF for
the Galactic center (left) and IMBH
(right) model. Binaries can exist at much higher energies for our "realistic" IMBH model compared to the "realistic" GC model. The reason for the difference is that for real systems, the stellar radius is an additional length scale of importance that starts to play a role. Binaries cannot be tighter than their own radius, and in this model we therefore allowed for much harder binaries in the IMBH model ($\xi<400$ for IMBH compared to $\xi<7.3$ for GC, see table \ref{t:t1}). If instead one would consider the evolution of binaries of compact remnants, their radii, or the radius where gravitational wave emission becomes important, are so small that there would not be a difference between the IMBH and GC models \citep[see also][for the role of stellar black holes in galactic nuclei]{OLe08}.

\begin{figure*}[!htb]
    \begin{center}
         \begin{tabular}{c c}
              \scalebox{0.35}{\includegraphics[angle=270]{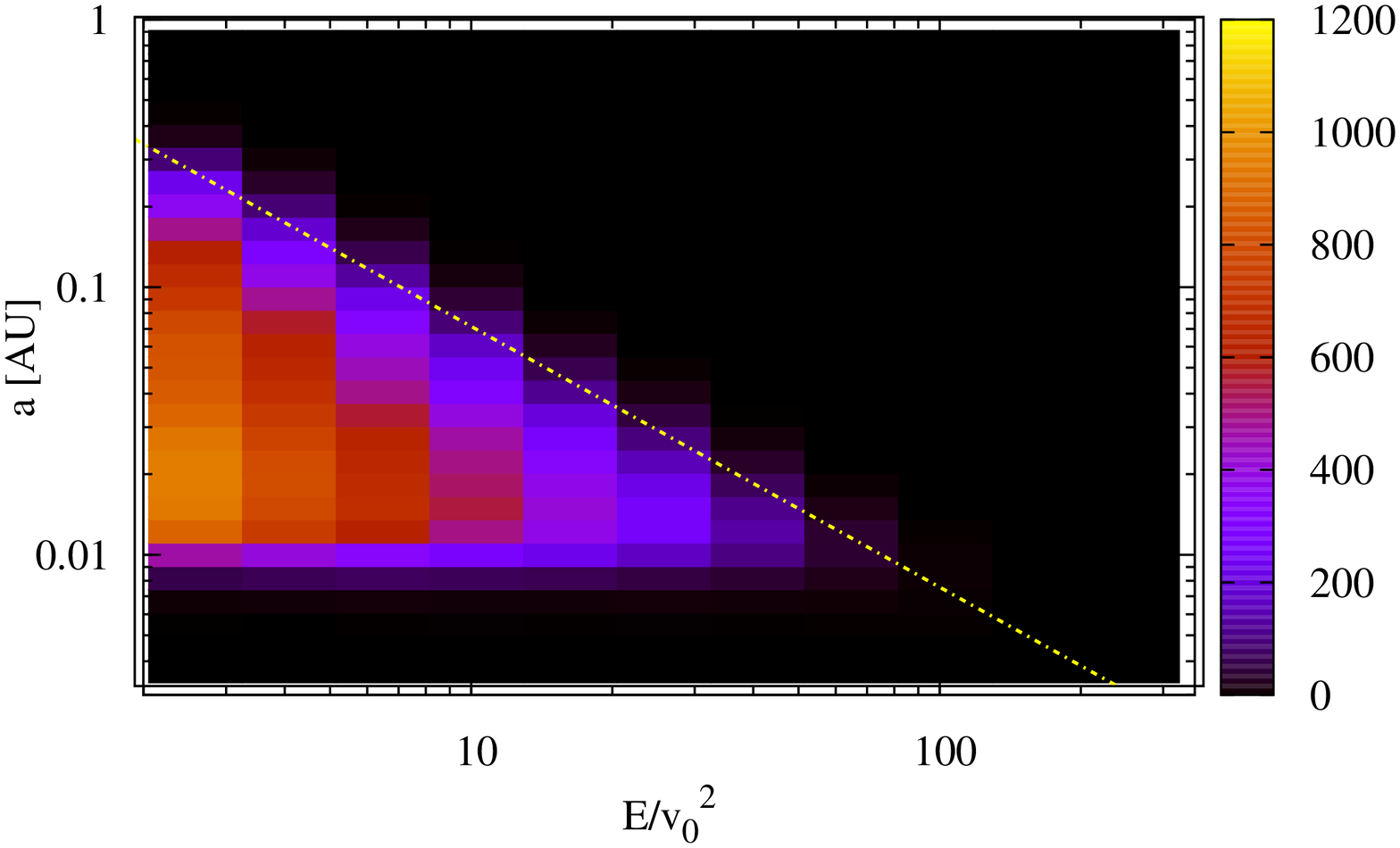}} &
              \scalebox{0.35}{\includegraphics[angle=270]{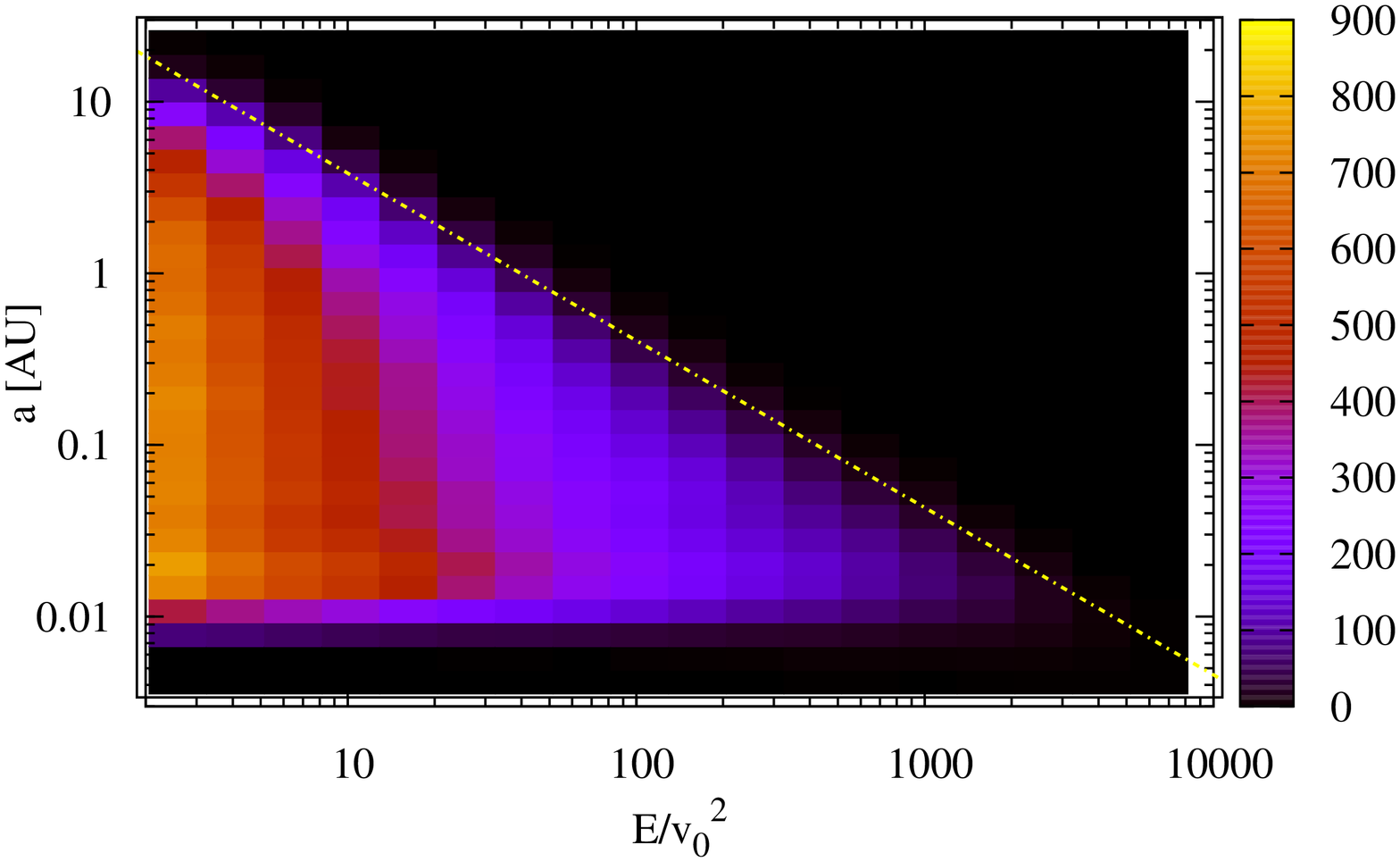}} \\
              \small \emph{(a)} & \small \emph{(b)}\\
         \end{tabular}
      \caption{Two dimensional distribution $E^{5/4}d^2N(E, a)/d\ln E d\ln a$ of binaries with arbitrary scaling for the GC model (figure a) and the IMBH model (figure b). The smaller the semi-major axis of the binary, the more resilient it is against ionization, and hence the larger the energy with respect to the MBH for which it can still exist. The line indicates the boundary $E=w_sGm/(2a)=4Gm/a$ where binaries approximately become soft. Since binaries are harder for the IMBH model, they can exist at much larger radii. Note that the horizontal axis of the figures have different scales.\label{f:Ea}}
    \end{center}
\end{figure*}

Another way of viewing the effect of ionization on the binary DF is shown in figure (\ref{f:Ea}). In this figure, the two dimensional distribution $E^{5/4}d^2N(E, a)/d\ln E d\ln a$ is displayed; the factor $E^{5/4}$ was added to factor out some of the energy dependence of the DF (see equation \ref{e:Nanalyt}). Again it is clear that at small energies binaries of all types can exist, whereas at high energies there is an $a$-dependent maximum energy that is approximately given by $E_s(a)=w_sGm/(2a)=4Gm/a$. Here $w_s=8$ is the same parameter is used in the model in \S\ref{ss:param}

\subsection{The binary fraction}\label{ss:fb}

\begin{figure*}[!htb]
    \begin{center}
         \begin{tabular}{c c}
              \scalebox{0.35}{\includegraphics[angle=270]{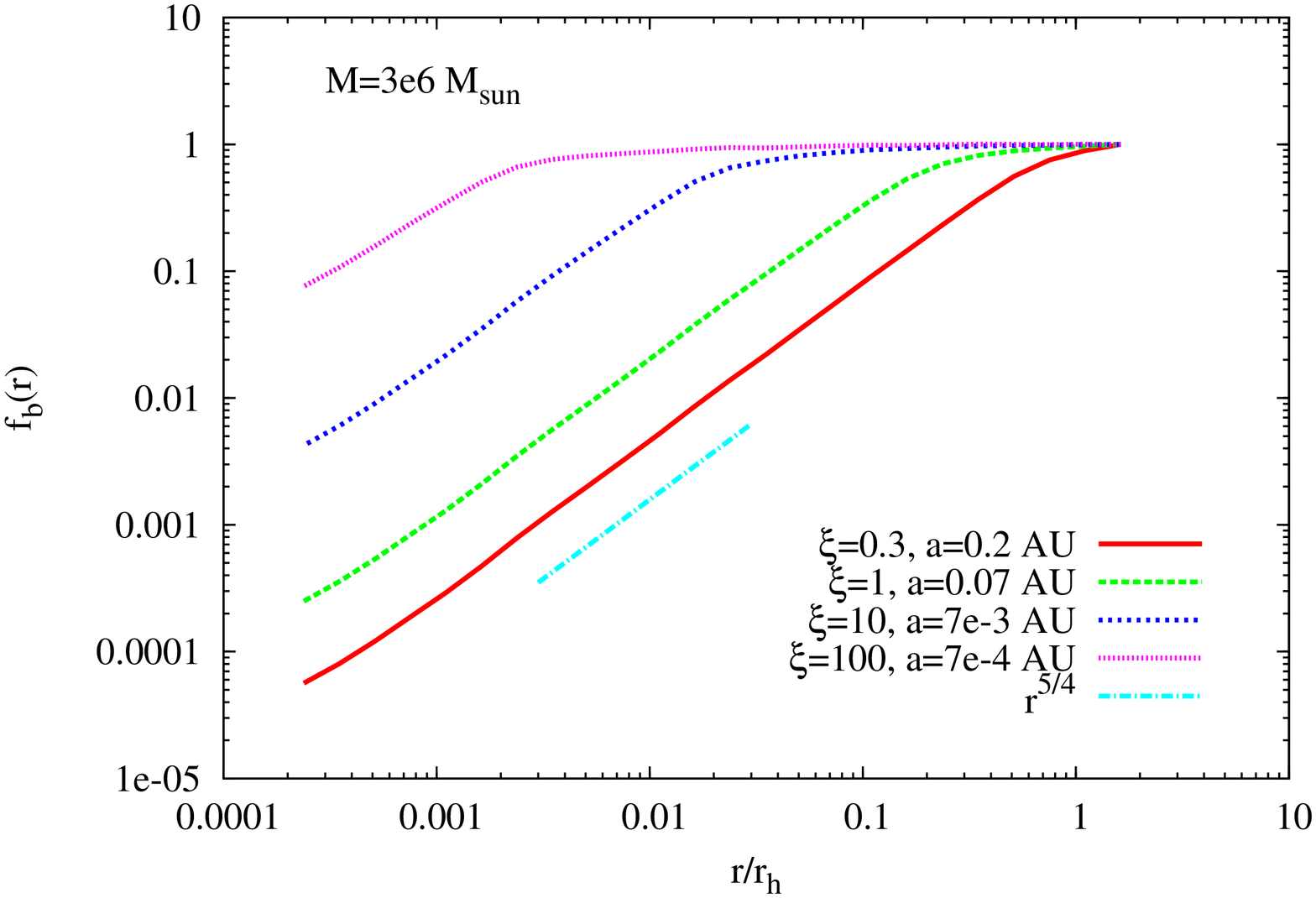}} &
              \scalebox{0.35}{\includegraphics[angle=270]{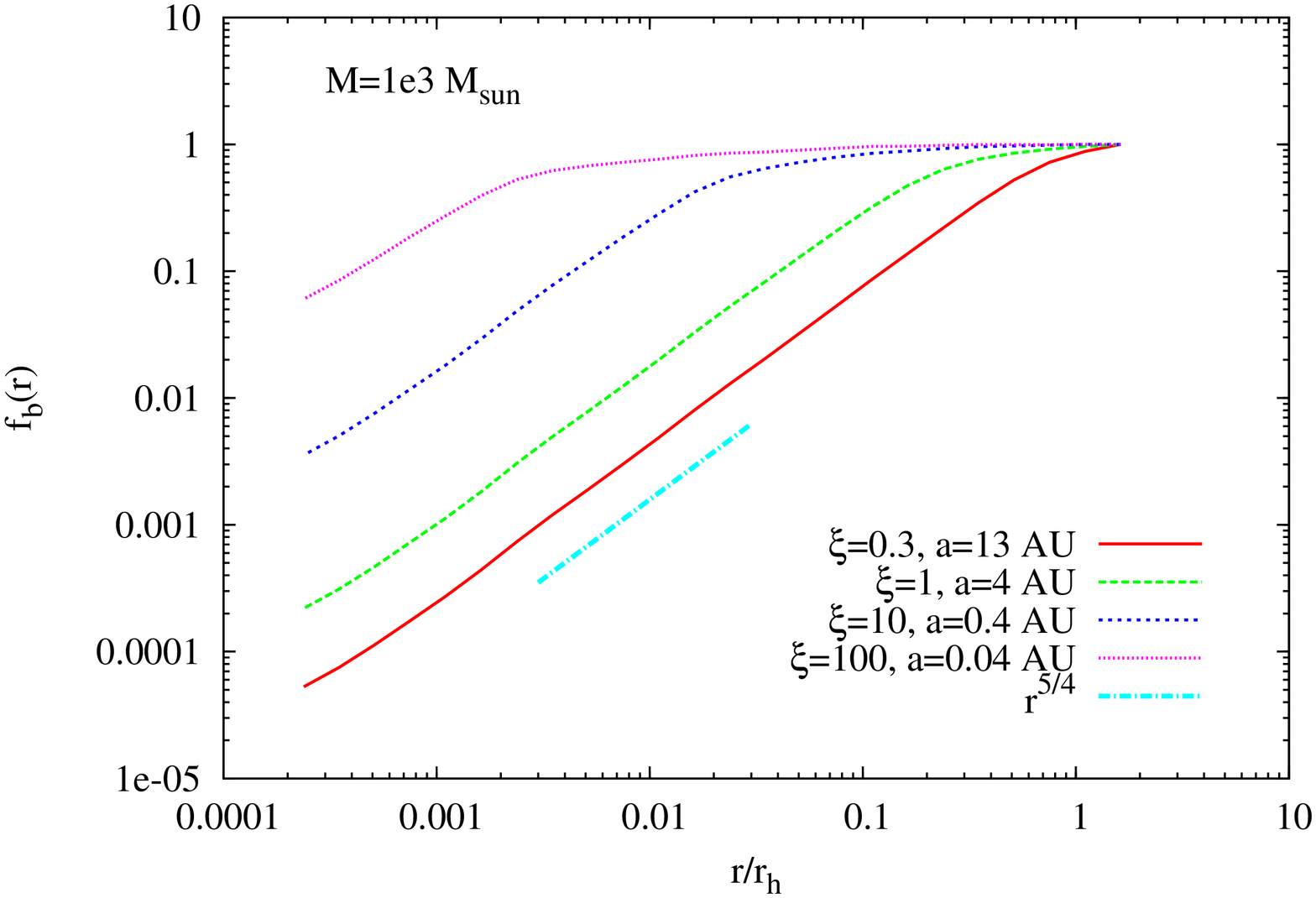}} \\
              \small \emph{(a)} & \small \emph{(b)}\\
              \scalebox{0.35}{\includegraphics[angle=270]{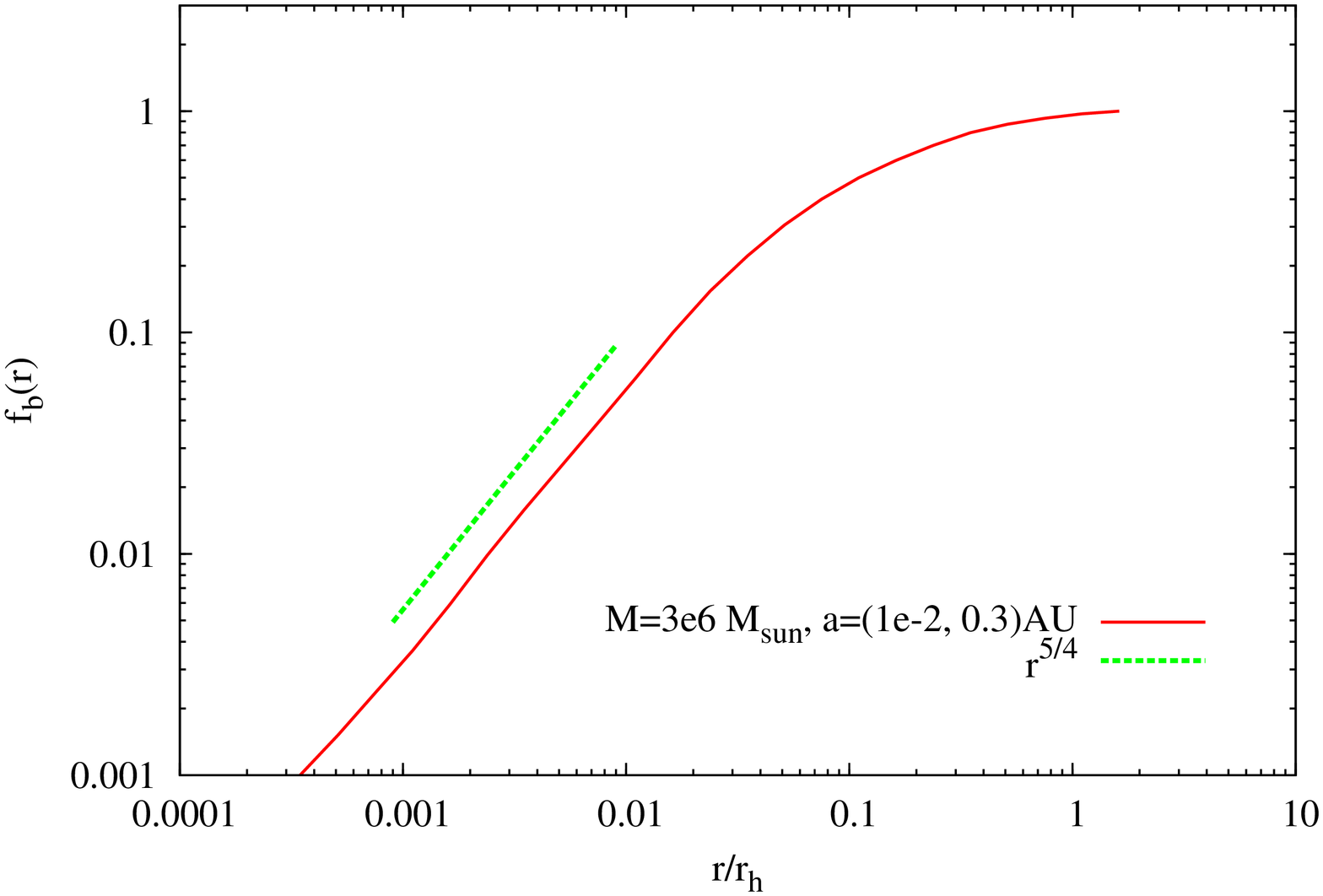}} &
              \scalebox{0.35}{\includegraphics[angle=270]{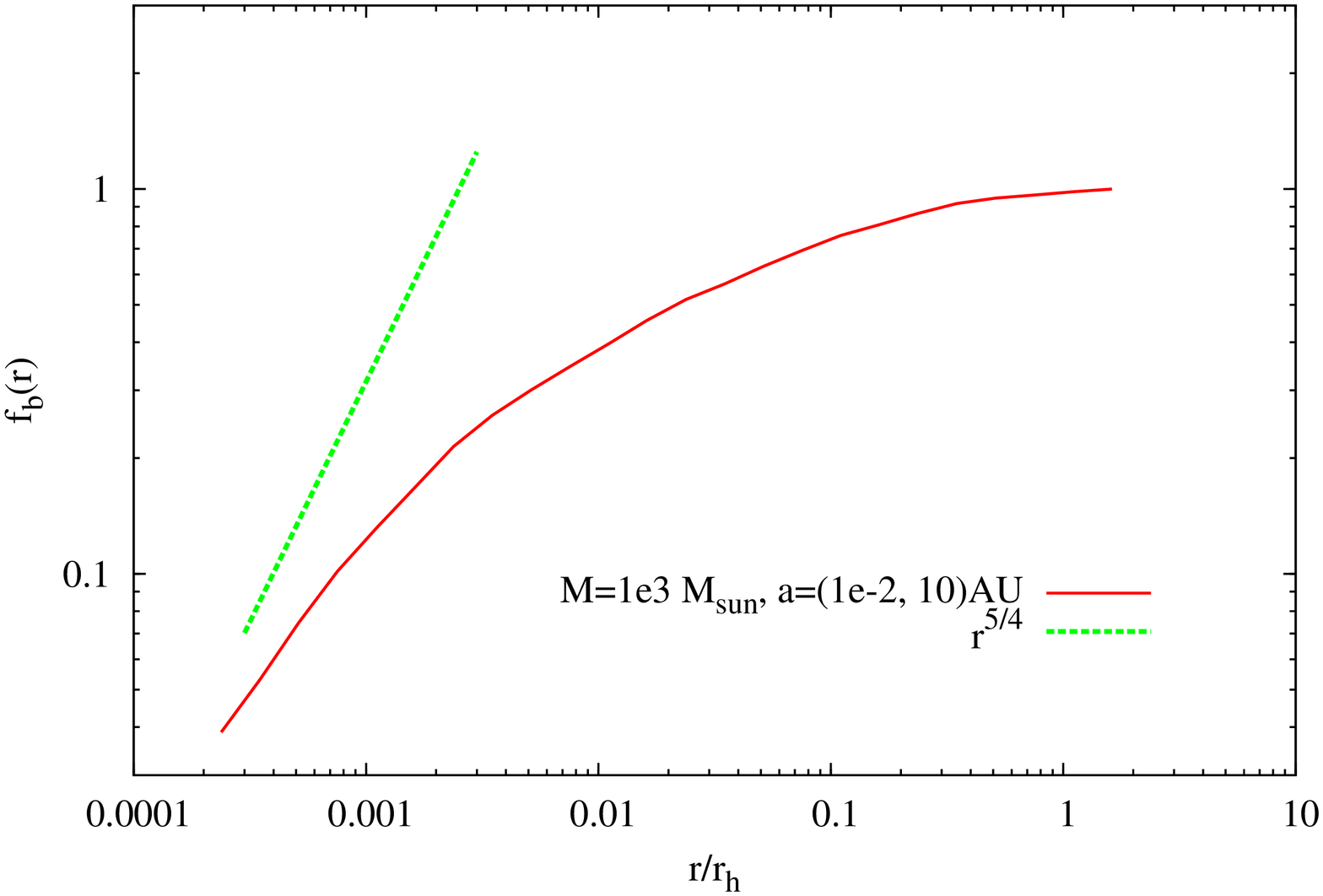}}\\
              \small \emph{(c)} & \small \emph{(d)}\\
         \end{tabular}
      \caption{Binary fraction as a function of radius. Figure ({\it
      a}) shows models II, figure ({\it b}) shows models III, figure
      ({\it c}) shows model IV, and figure ({\it d}) shows model V. As
      in figure (\ref{f:fE}), it is clear that hard binaries can
      survive closer to the MBH. The straight line is $\propto r^{5/4}$,
      which is the radial dependence of the binary fraction if only highly eccentric orbits
      contribute to the density profile (see equation
      \ref{e:fbflyby}). Note that these binary fractions are scaled such that $f_b(r=r_h)\equiv1$; if the binary fraction of the type considered is lower than 1, as will typically be the case, the actual binary fraction needs to be multiplied by the required factor. In particular, for the Galactic center the actual binary fraction will be smaller by a factor $\lesssim0.1$. In the IMBH model, binaries are present even at very small radii, and the $r^{5/4}$ limit is not reached.\label{f:fb}}
    \end{center}
\end{figure*}

The number density of stars is related to the DF by equation
(\ref{e:nr}). Let the binary fraction $f_b(r)$ be defined as the fraction of the
number of binaries in a particular scenario, compared to the number of
binaries in the absence of a loss-cone and three-body interactions,

\begin{equation}\label{e:fb}
f_b(r)\equiv{n_b(r)\over n_{p}(r) } = {\int d^3v f_{b}(E, J)\over \int d^3v f_{p}(E)}
\end{equation}
(the subscript $p$ stands for ``point particles''). The actual binary
fraction is much smaller than this, and equals the fraction of
binaries at the radius of influence $f_{b, 0}$, times
$f_{b}(r)$. Since we did not attempt to model the evolution of binaries
outside the radius of influence, we keep $f_{b, 0}$ as a free parameter (so $f_{b}(r=r_h)=1$). 

Even if no binaries at all are bound to the MBH beyond a certain
energy, there is always a finite density at any distance from the MBH
because of loosely bound, highly eccentric orbits. The density of such
orbits in a Kepler potential increases as $n\propto r^{-1/2}$ towards
the MBH. Since the density of single stars is $n_{s}\propto r^{-7/4}$,
the binary fraction due to highly eccentric orbits is 

\begin{equation}\label{e:fbflyby}
f_{b}^{\rm fly-by}(r)\propto r^{5/4}.
\end{equation}

Figure (\ref{f:fb}) displays the binary fraction $f_b(r)$ as a
function of radius, for models II --- V. As
in figure (\ref{f:fE}), one can see that the harder the binary, the
higher the binary fraction close to the MBH. Again the results for binaries of a given hardness are very similar when the IMBH and GC cases are compared, for the same reason that led to the similarity between the DFs. For the realistic GC model, the binary fraction can be
substantial outside $\sim10\%$ of the radius of influence. Closer to
the MBH, the profile drops as $f_b(r)\propto r^{5/4}$, implying that
only ``fly-by binaries'' contribute to the density profile there. In contrast, binaries can exist in substantial amounts in all regions near an IMBH, the reason being (again) that much harder binaries are allowed for that case. 

\subsection{Binary disruption rates}\label{ss:disr}

\begin{figure*}[tb]
    \begin{center}
         \begin{tabular}{c c}
              \scalebox{0.35}{\includegraphics[angle=270]{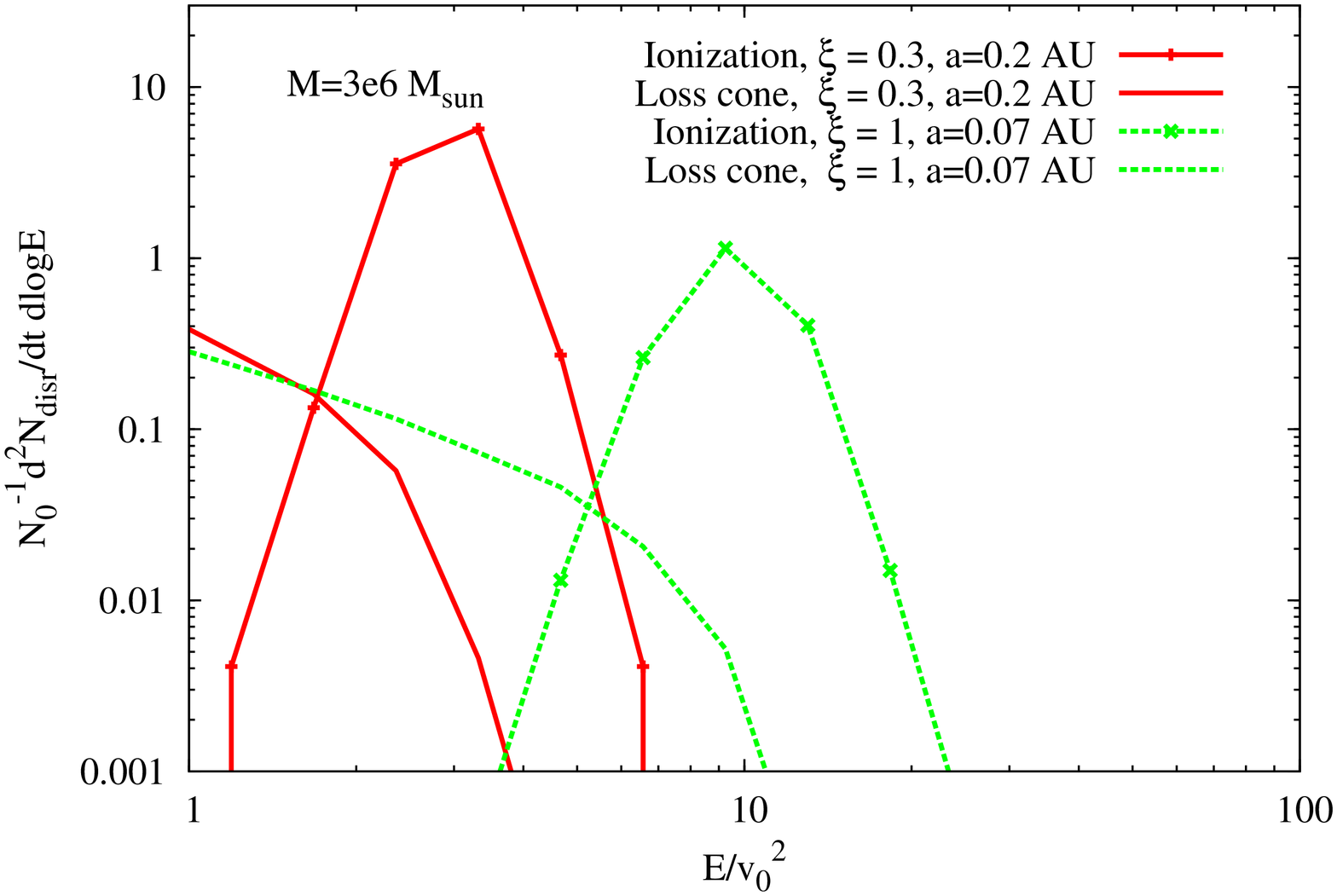}} &
              \scalebox{0.35}{\includegraphics[angle=270]{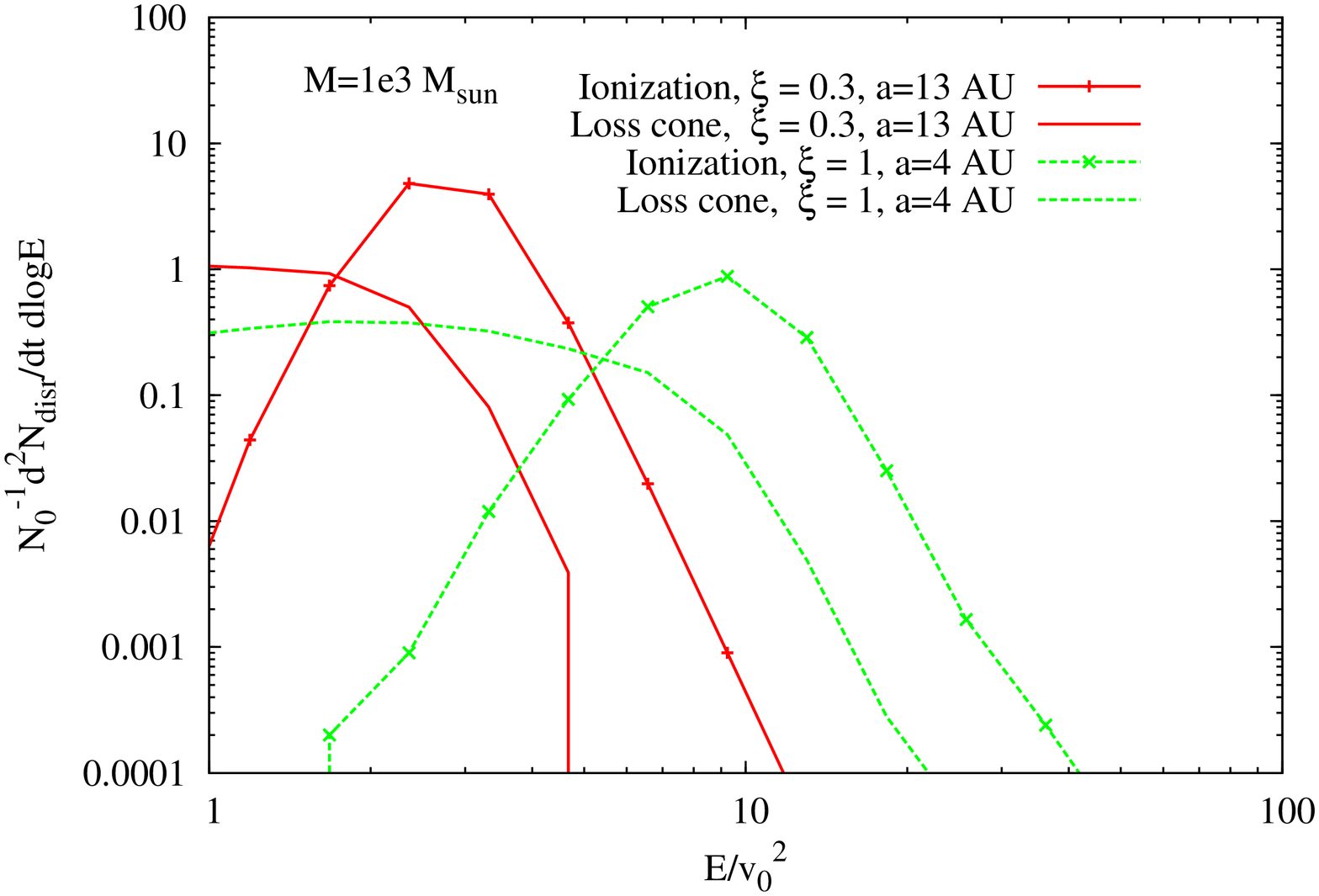}} \\
              \small \emph{(a)} & \small \emph{(b)}\\
              \scalebox{0.35}{\includegraphics[angle=270]{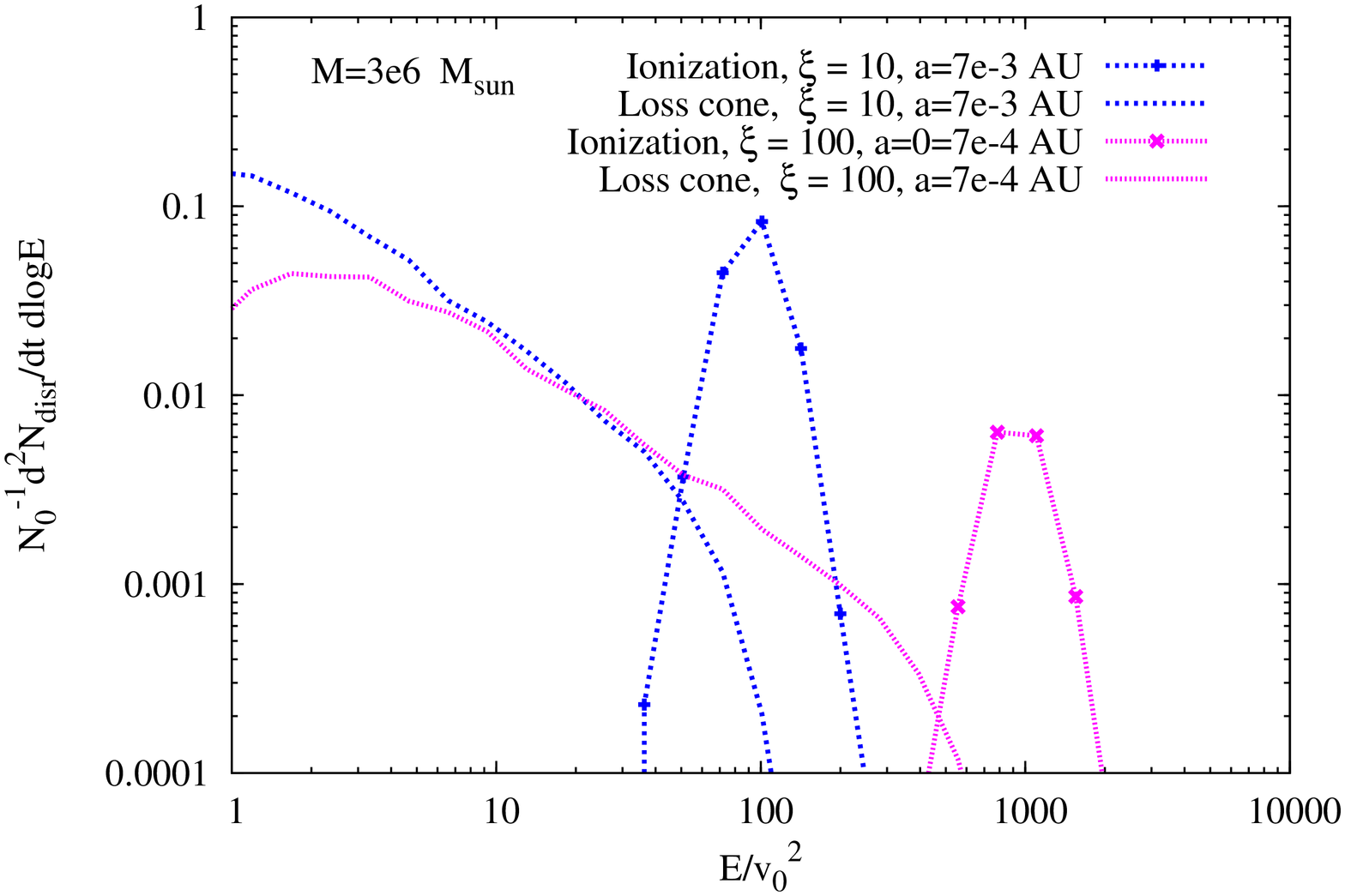}} &
              \scalebox{0.35}{\includegraphics[angle=270]{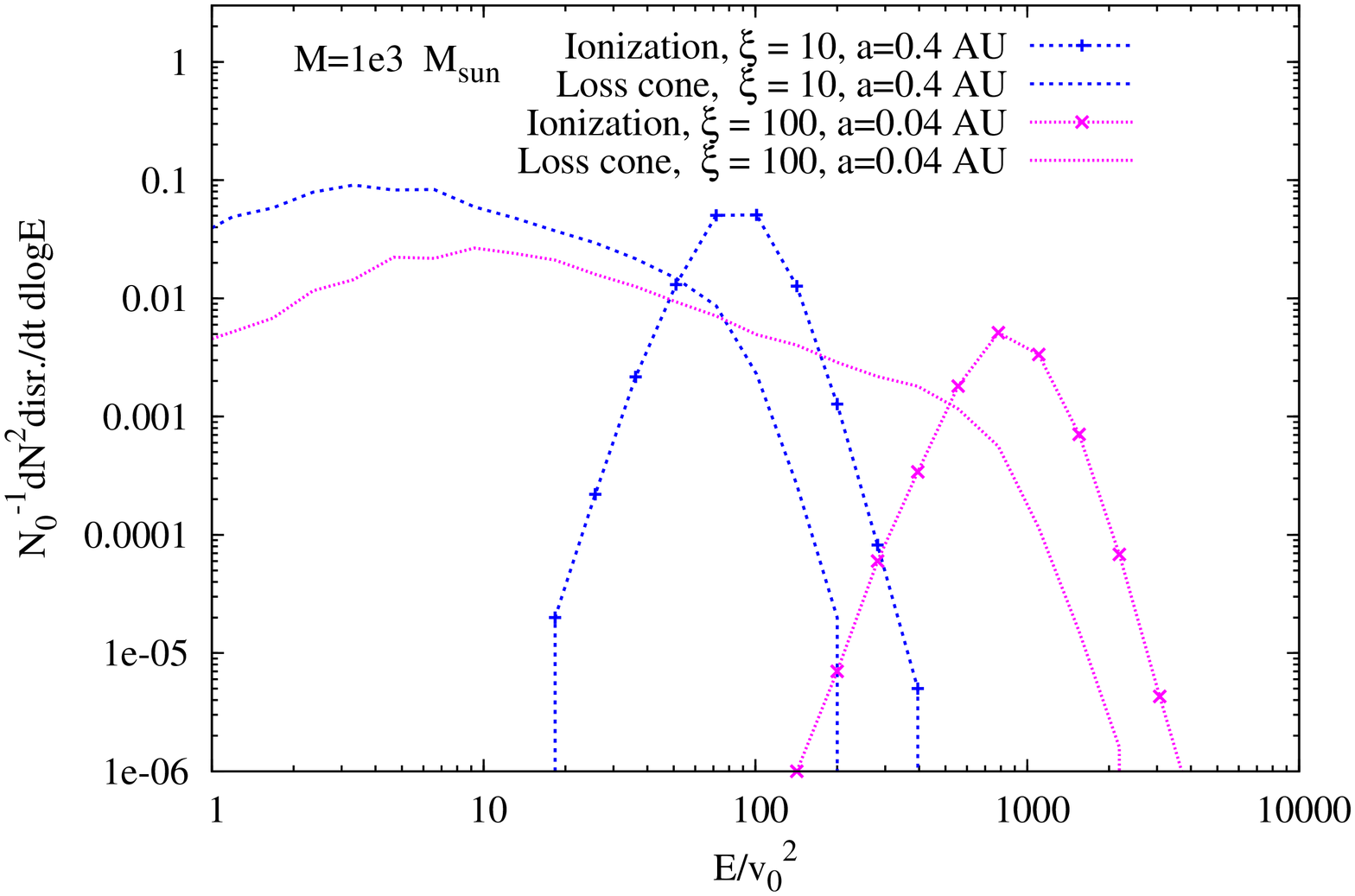}}\\
              \small \emph{(c)} & \small \emph{(d)}\\
              \scalebox{0.35}{\includegraphics[angle=270]{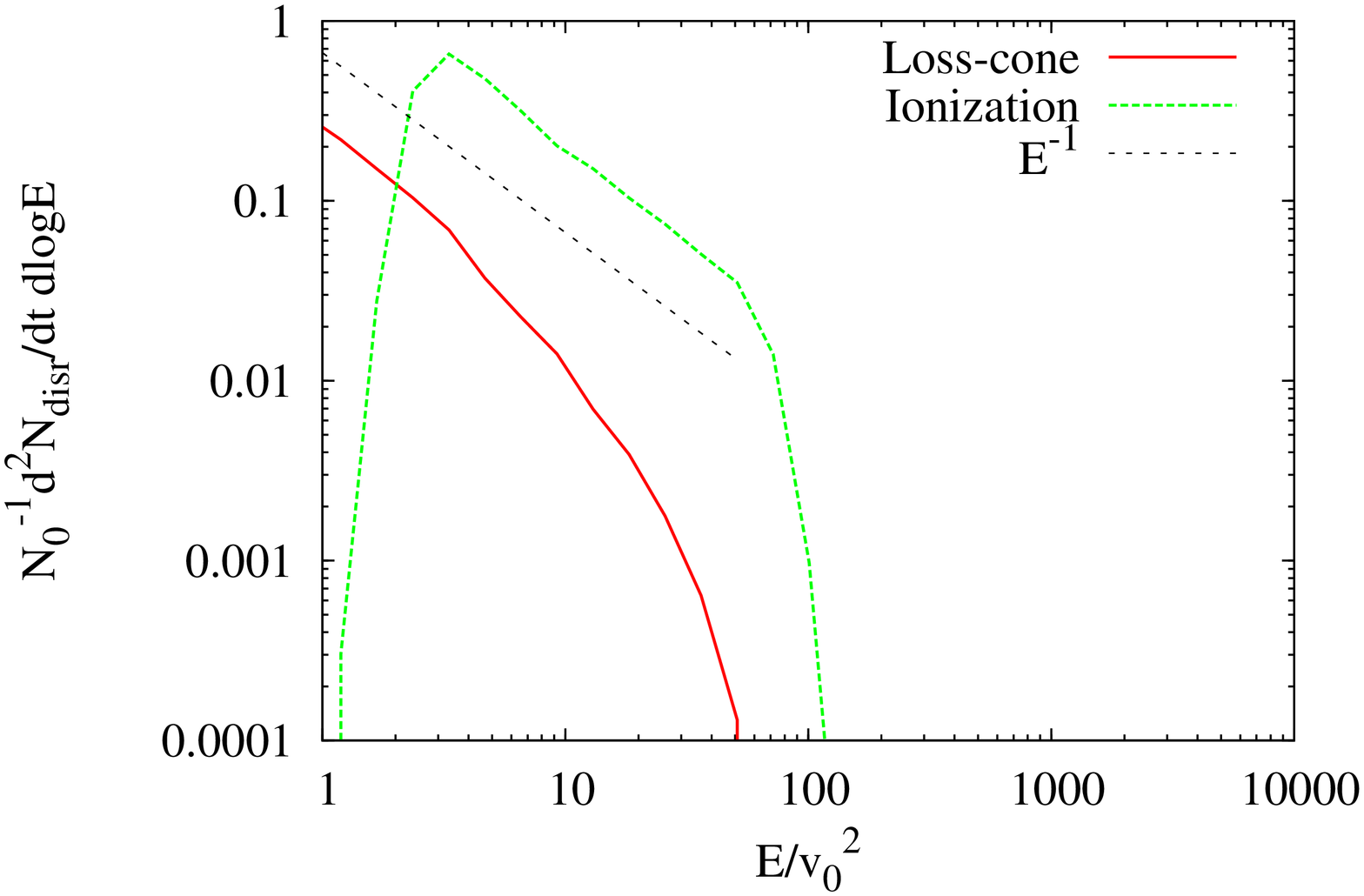}} &
              \scalebox{0.35}{\includegraphics[angle=270]{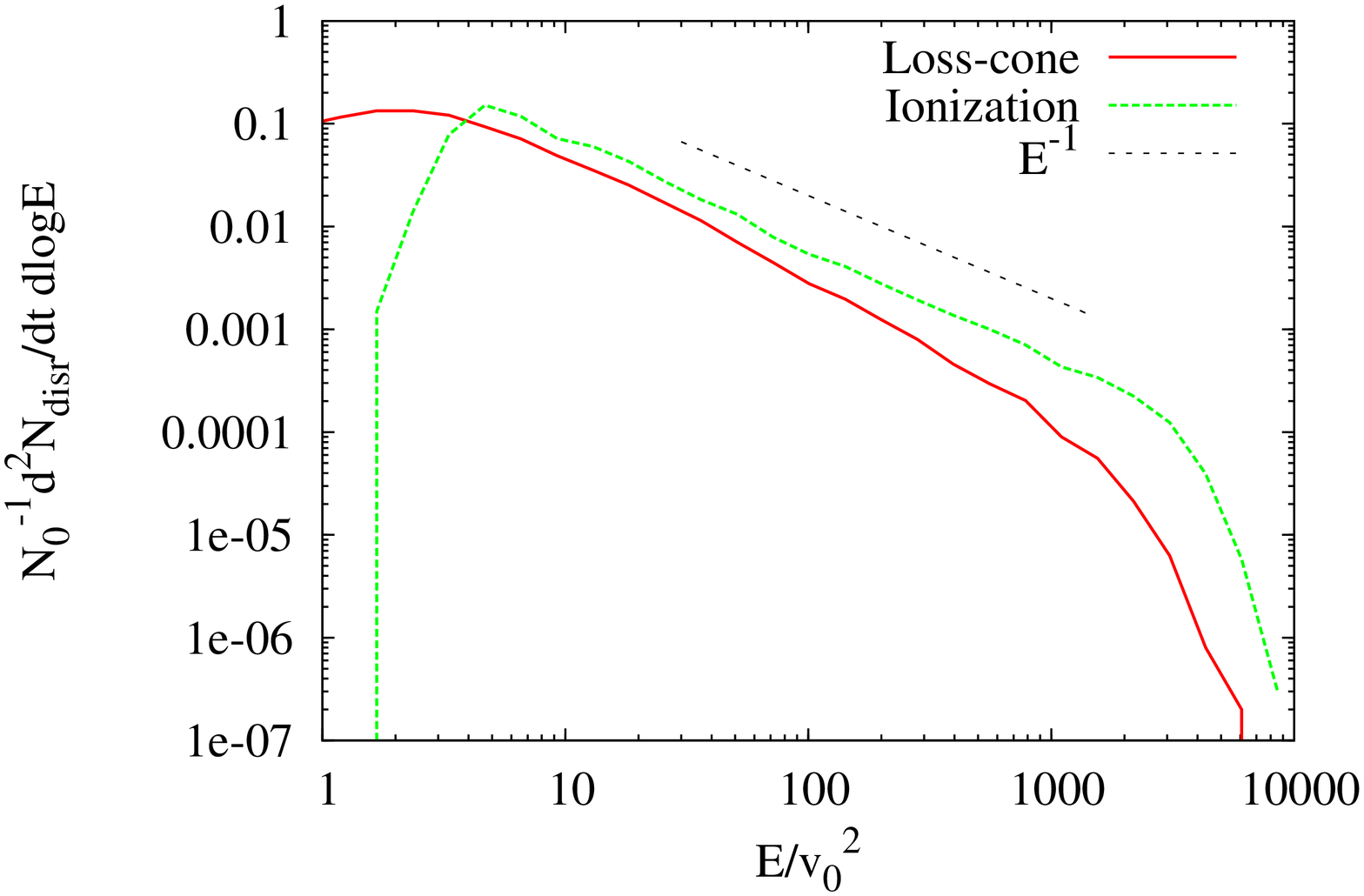}} \\
              \small \emph{(e)} & \small \emph{(f)}\\
         \end{tabular}
      \caption{Disruption rate per unit $\ln E$, per binary that is present in the system. A distinction is made between loss-cone
      disruptions and ionizations. Figures ({\it a, c, e}) are for a
      $\Mbh=3\times10^{6}$ MBH, figures ({\it b, d, f}) are for an
      IMBH. Figures ({\it e}) and ({\it f}) are the ``realistic''
      models, the others are cases where all the binaries start at a
      given initial semi-major axis, as indicated in the legends. Note that the top figures have a different scale. For all but the softest binaries, tidal disruptions dominate over ionizations at low energies. As the binaries diffuse to higher energies, they become softer with respect to their environment, until they reach the point where their internal energy becomes smaller than the kinetic energy of the stars around them, at which point they are rapidly ionized, and ionization rates exceed tidal disruption rates at sufficiently high energies. This process is so rapid that it depletes all binaries and at higher energies, binaries can no longer exist; the ionization rate is therefore localized in energy. The realistic models give tidal disruption rates of $7\times10^{-7}(f_{b, 0}/0.05)\peryr$ for the GC and $10^{-5}(f_{b, 0}/0.1)\peryr$ for IMBHs, where $f_b$ is the binary fraction of stars far away from the MBH. See text for discussion of these rates.\label{f:disr}}
    \end{center}
\end{figure*}

Figure (\ref{f:disr}) shows the rate $N_{0}^{-1}\left(d^2N/d\ln E
dt\right)$ at which binaries are destroyed by ionization or tidal
disruptions for the different models. For soft
binaries $(\xi=0.3)$, the ionization rate quickly becomes higher than the
loss-cone rate as energy increases. For binaries that are hard when they are marginally
bound to the MBH, loss-cone effects dominate at low energies. These binaries can only ionize on very eccentric orbits where they spend a small fraction of time in very hot regions near the MBH. At higher energies,
ionizations always dominate over loss-cone disruptions. Since the
time-scale for ionizations is much smaller than the relaxation time,
energy diffusion is not sufficiently rapid to replace binaries that
have been destroyed, and the stars are depleted at high $E$. Ionizations for binaries of a given initial semi-major axis is thus localized around a small range in energies, while tidal disruptions decrease gradually as the energy grows, and then suddenly drops off because ionization effects annihilate all binaries rapidly. The analytical toy model captures these aspects of the disruption rates.

The situations for $\Mbh=10^3\Mo$ and $\Mbh=3\times10^6\Mo$ are quite similar. The only place where the difference in mass comes into play is that for binaries of a given hardness $\xi$, the point where the full loss cone becomes empty is located at a higher energy for MBHs of lower mass. The turn-over from the full to empty regime is most clear for the case where $\Mbh=10^3\Mo, \xi=100$ (model IIId).

In contrast, the difference between the realistic IMBH and GC models is again quite clear. As argued before, very hard binaries cannot exist near the GC MBH, because of the finite size of the stellar radius. Loss-cone rates do dominate for very low energies, but ionization effects take over quickly as the energy increases, and at an energy of $E\sim100v_0^2$, most binaries have been ionized. For IMBHs, the majority of the binaries are so hard that loss-cone effects dominate everywhere.

In the GC model presented here, the tidal disruption rate of binaries
in the Galactic center is $7\times10^{-7}(f_{b, 0}/0.05)\peryr$, where $f_{b, 0}$ is the binary fraction of stars far away from the MBH. This rate is lower by an order of magnitude than the rate found in \citet{YuQ03}. The cause of the difference is that we only considered binaries that are bound to the MBH. Figure (\ref{f:disr}e) shows that the tidal disruption rate continues to increase as the energy decreases, since the loss-cone in this regime is still empty for binaries. The tidal disruption rate of binaries can even be much larger than that given in \citet{YuQ03} if there are massive perturbers at a few pc from the MBH, or if the potential is significantly triaxial at those distances \citep{Per07}.

For the IMBH model, we find a tidal disruption rate of $10^{-5}(f_{b, 0}/0.1)\peryr$. In contrast to the GC model, this rate will not increase much if binaries unbound to the MBH are considered, since most binaries are in the full loss-cone regime. The rate we find here is much higher than the rate estimated in \citet{Pfa05}, who found an event rate of $\sim10^{-7}(f_{b, 0}/0.1)\peryr$. The discrepancy can be traced to the fact that the models for the ambient cluster were quite different, with our model having a much smaller relaxation time. If we were to use the relaxation time of \citet{Pfa05}, we find an event rate of $2\times10^{-7}(f_{b, 0}/0.1)\peryr$, which is in good agreement with his results.

It is evident that the disruption rates near IMBHs are highly uncertain, given that there is not much known about such systems. It is interesting to note, however, that current models of tidal disruptions, both of binaries and of single stars, typically lead to event rates that are so high that the IMBH disrupts more than its own mass within a Hubble time \citep[e.g.][]{Wan04, Hop09}. It is therefore unlikely that IMBH systems can exist in this form for a Hubble time: either the IMBH mass will grow, or the system around the MBH will expand and become less dense \citep{Mar80}, leading to lower event rates; this effect is seen in $N$-body simulations \citep{Tre07}. An obvious third possibility is that IMBHs do not exist.

\begin{figure}[!h]
\includegraphics[angle=270,scale=.37]{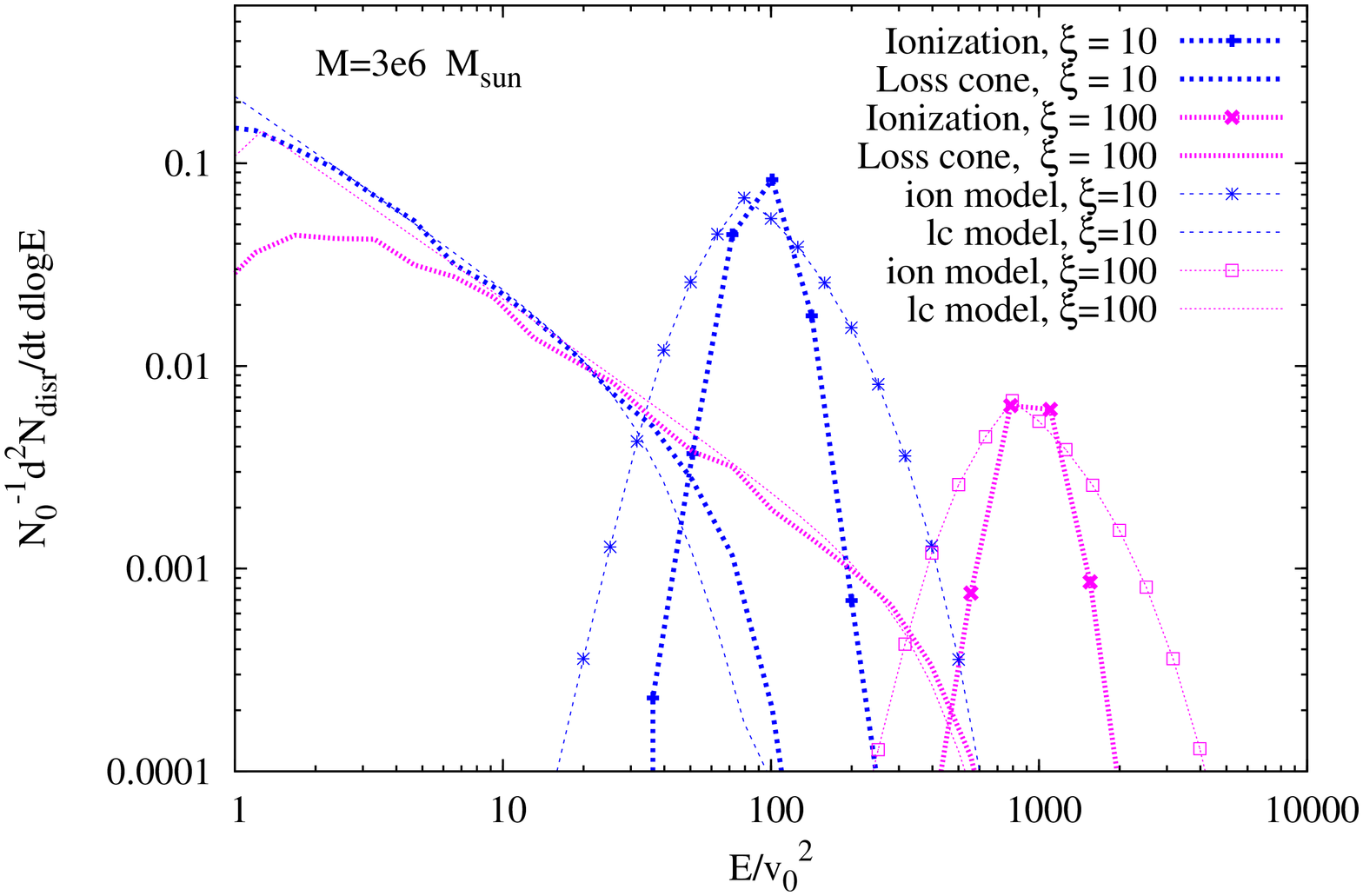}
\figcaption{Disruption rates for the $\Mbh=3\times10^6$, $\xi=(10, 100)$ models (thick lines), compared with the analytical approximation (thin lines with same colors for given model). \label{f:andisr}}
\end{figure}

In figure (\ref{f:andisr}) we show the same curves as in figure (\ref{f:disr}c), and add the analytical approximations from \S\ref{ss:param}. The qualitative agreement is good, and the analytical approximation captures the main dynamical effects, such as the localized peaks in ionization, the energy dependence and magnitude of the tidal disruption rates, and the cut-off of all rates at high energies. We therefore believe that the analytical arguments presented in \S\ref{ss:param} are essentially correct.

\citet{Tre07} report that tidal disruption rates of binaries were much lower in their $N$-body simulations than analytical predictions suggest. The binaries in \citet{Tre07} were hard compared to the average kinetic energy in the cluster, but presumably many binaries were soft compared to the environment near the radius of influence, where the velocity dispersion is higher than average. Figure (\ref{f:disr}b) shows that for soft or marginally hard binaries, the ionization rate is larger by an order of magnitude than the tidal disruption rate, similar to what \citet{Tre07} found. This implies that the loss-cone rate is smaller by a factor $\gamma_{lc}/(\gamma_{lc}+\gamma_{a})<0.1$ than one would expect if ionizations are not accounted for (see equation [\ref{e:tiddisrrate}]). This factor brings the analytical model in qualitative agreement with the results by \citet{Tre07}.

\subsection{Internal binary evolution: semi-major axis and exchange rate}\label{ss:bin}

\begin{figure*}[!htb]
    \begin{center}
         \begin{tabular}{c c}
              \scalebox{0.35}{\includegraphics[angle=270]{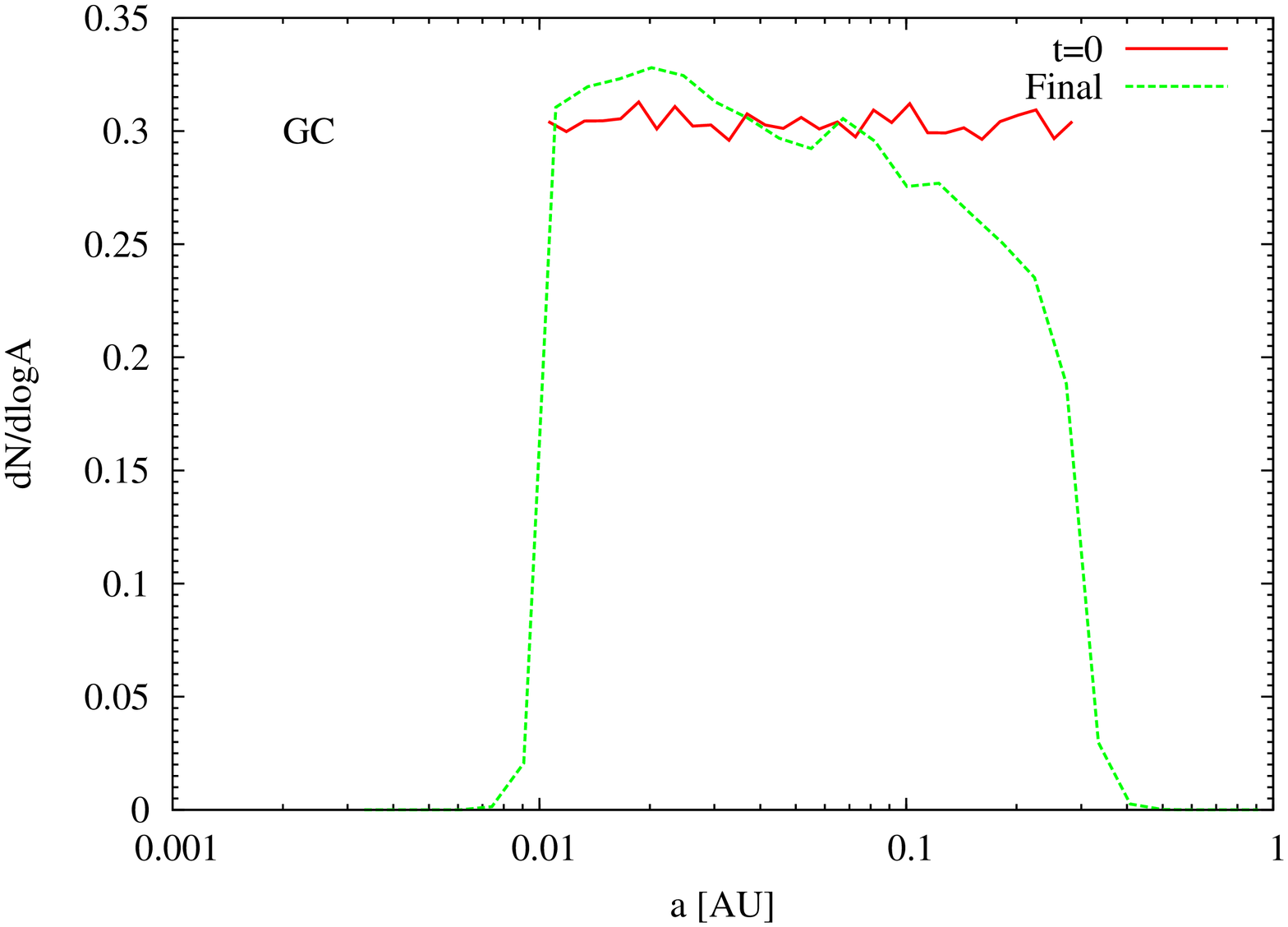}} &
              \scalebox{0.35}{\includegraphics[angle=270]{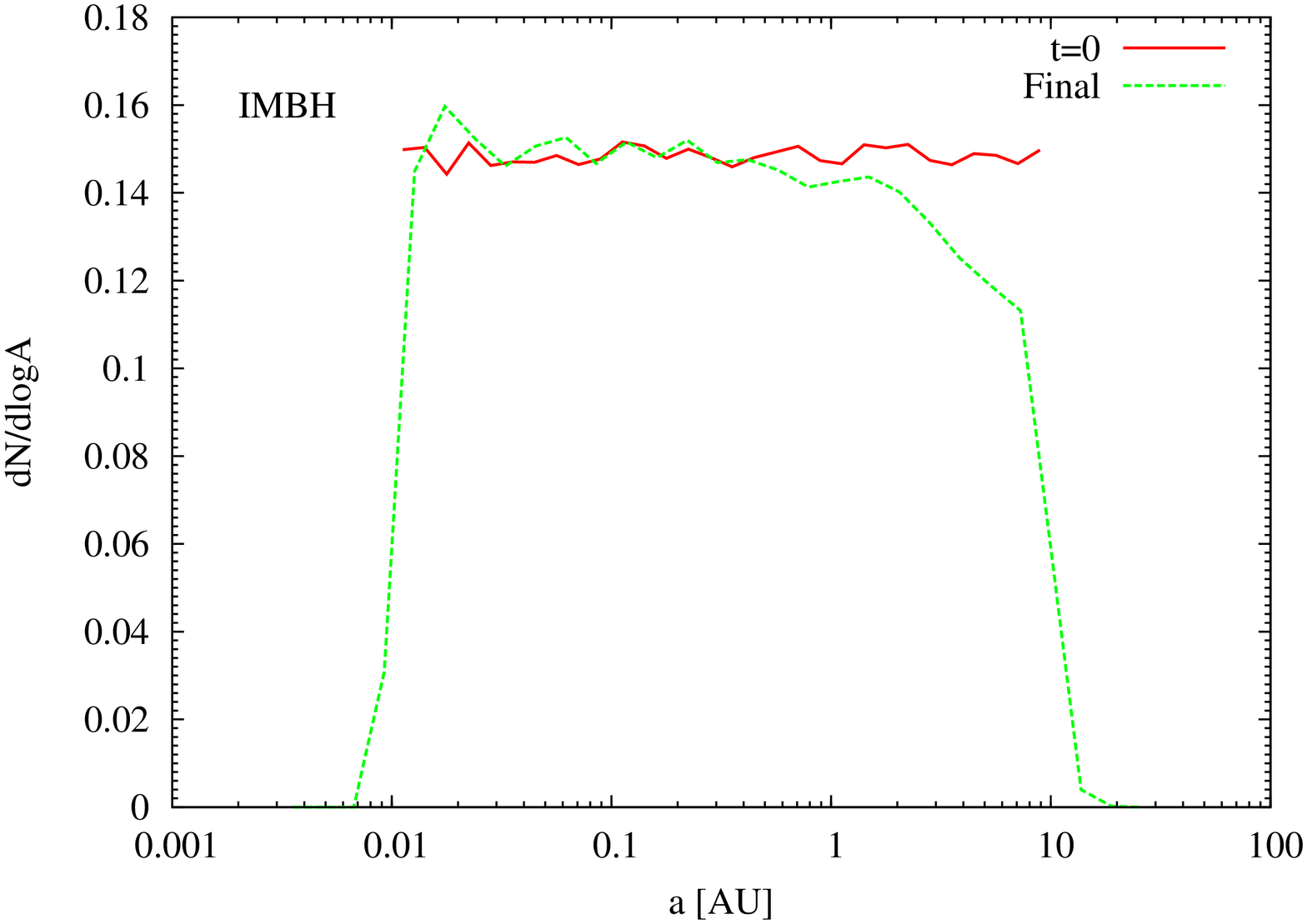}}\\
              \small \emph{(a)} & \small \emph{(b)}\\
         \end{tabular}
      \caption{The initial and final distribution per $\ln a$ of the binaries for
      the GC model ({\it a}) and the IMBH model ({\it b}).\label{f:a}}
    \end{center}
\end{figure*}

In figure (\ref{f:a}), the evolution of the distribution $dN/d\ln a$ of
semi-major axes are shown for times $t=0$ and
$t=T_{0}$ for the GC and IMBH models. There has been some hardening of binaries, and some of
the soft binaries are destroyed, leading to a moderate change in the
distribution. The relative change in the semi-major
axis is smaller for harder binaries, which have a smaller
cross-section. However, as discussed in \S \ref{s:times}, the
semi-major axis distribution cannot evolve by very much (assuming that
there were no very soft binaries at $E/v_{0}^2=1$), because the
diffusion and destruction rates are larger than the rate for
hardening. It is not very clear from the figure that ``Heggie's law'',
stating, roughly, that hard binaries become harder while soft binaries
become softer, is of relevance here. Hard binaries do not have time to
become harder, and when they move to higher energies, where the
densities are higher, they tend to become softer so that they are
disrupted. Soft binaries do become softer, but this happens so rapidly
that they are quickly ionized, so that they do not count in the final
statistics.

\begin{figure*}[!htb]
    \begin{center}
         \begin{tabular}{c c}
              \scalebox{0.35}{\includegraphics[angle=270]{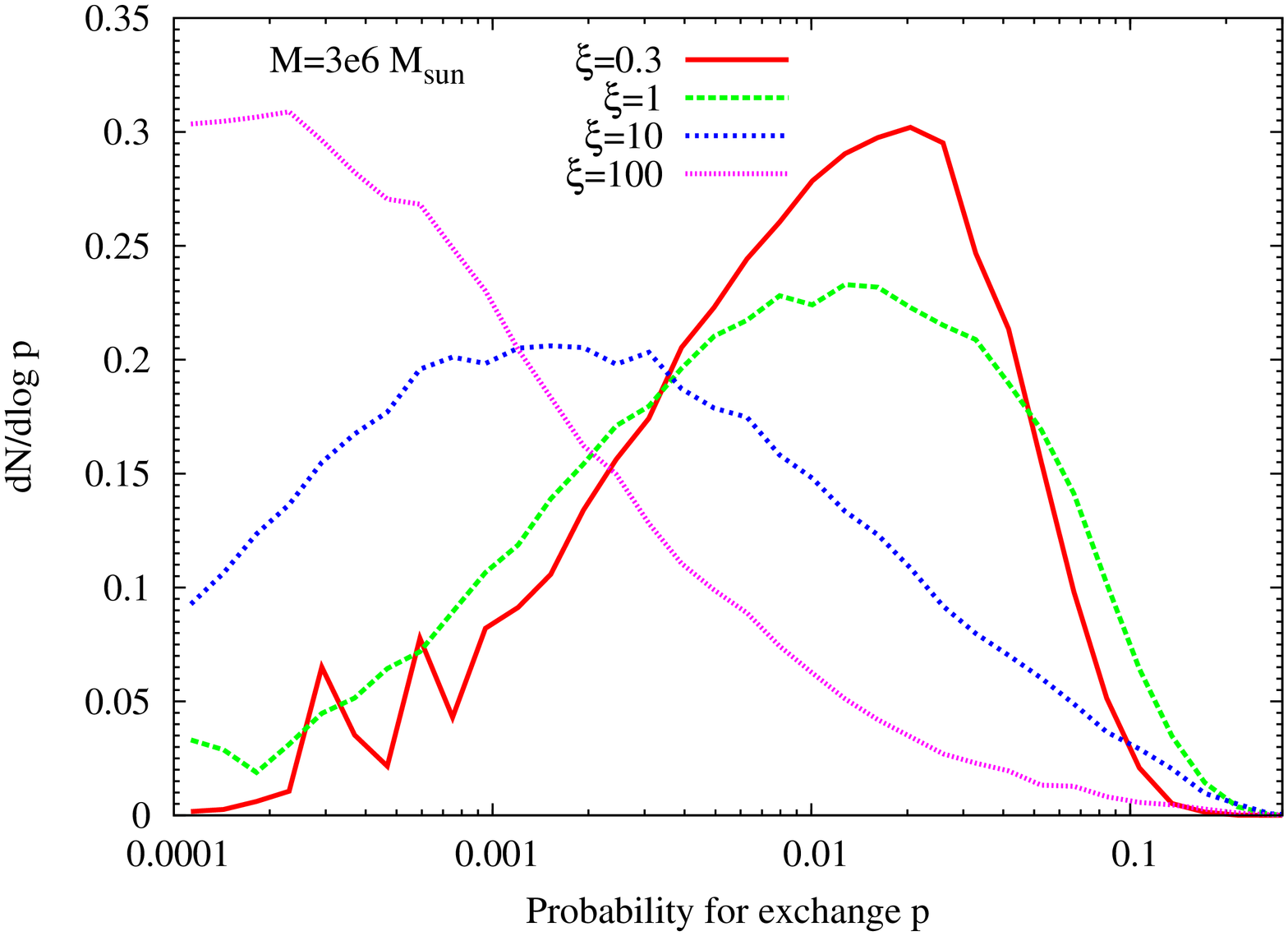}} &
              \scalebox{0.35}{\includegraphics[angle=270]{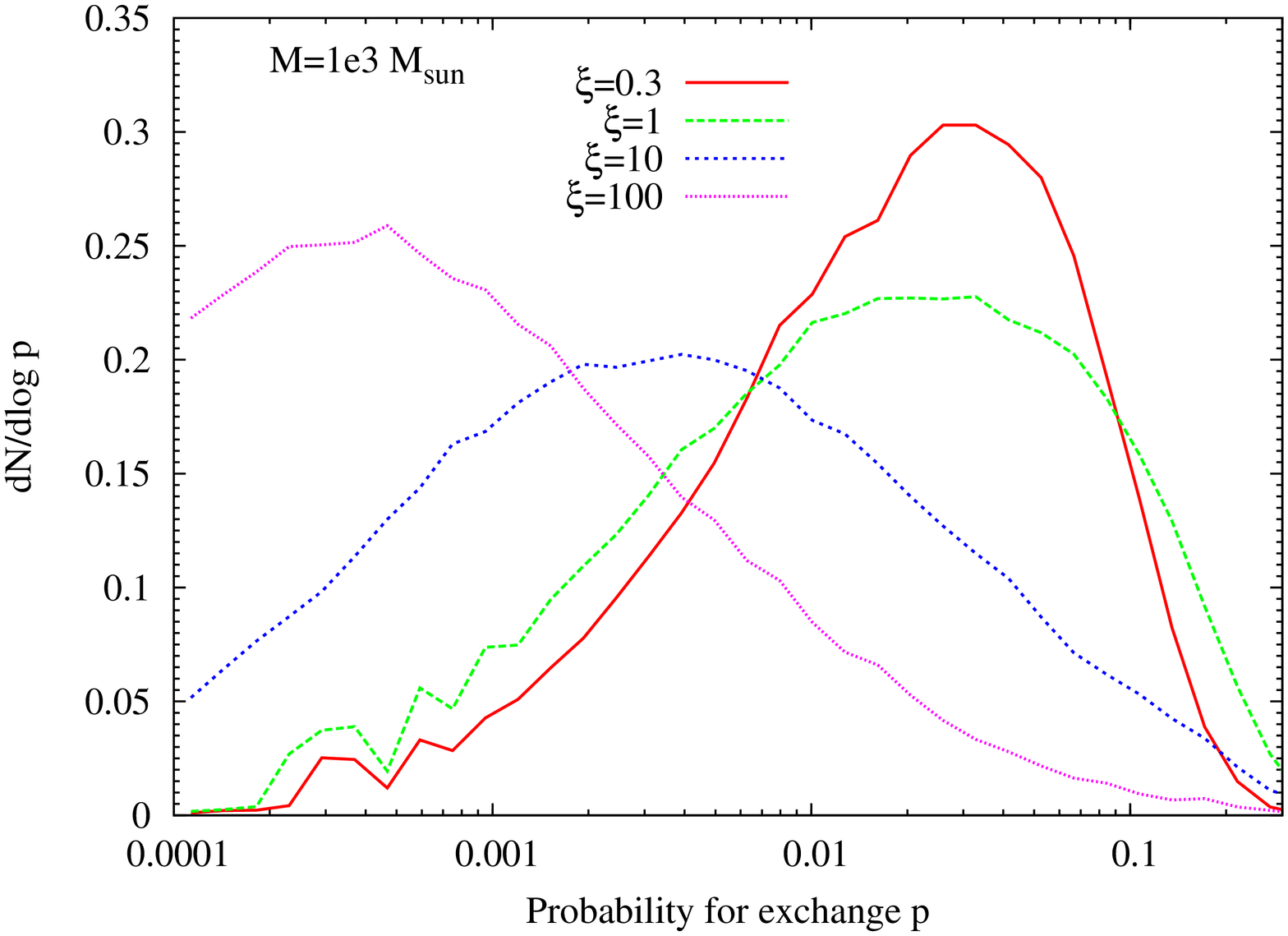}} \\
              \small \emph{(a)} & \small \emph{(b)}\\
              \scalebox{0.35}{\includegraphics[angle=270]{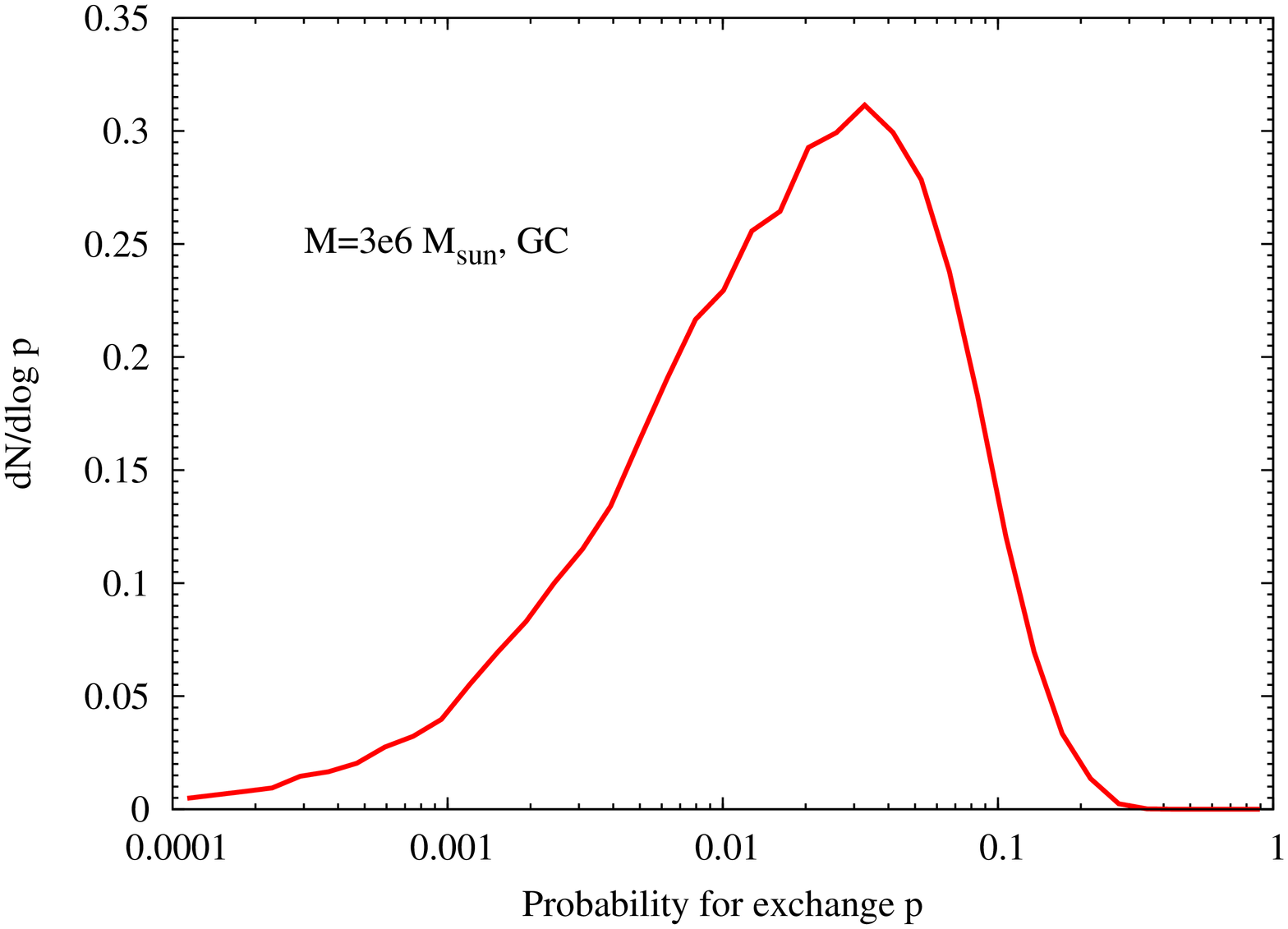}} &
              \scalebox{0.35}{\includegraphics[angle=270]{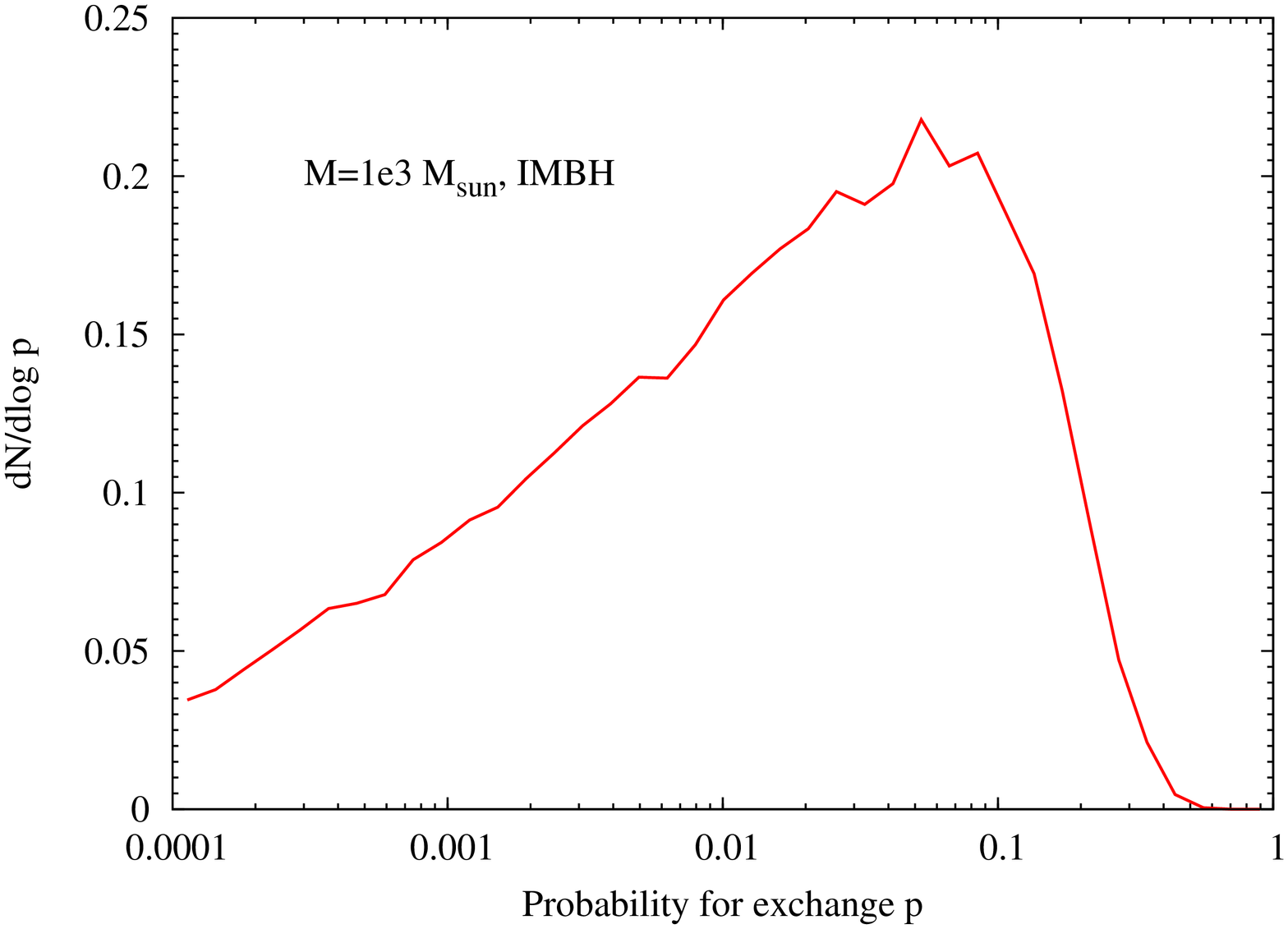}}\\
              \small \emph{(c)} & \small \emph{(d)}\\
         \end{tabular}
      \caption{Distribution of the log of the probability that a
      binary has experienced an exchange interaction (see equation
      \ref{e:Pex}). The probability is usually smaller than 0.1. For
      the Galactic center model, the mean exchange probability is
      $\langle p_{\rm ex}\rangle=0.03$,
      and for the IMBH model $\langle p_{\rm ex}\rangle=0.04$.
      \label{f:pex}}
    \end{center}
\end{figure*}

Figure (\ref{f:pex}) shows the distribution of probabilities that a
binary has undergone an exchange interaction for models (II ---
V). Binaries with $\xi=1$ at $E/v_0^2$ have the highest probability:
harder binaries have smaller cross-sections, while softer binaries are
more likely to be ionized and do not survive for a relaxation
time (this is in particular true for even softer binaries with $\xi<0.3$, not shown in this figure). Nevertheless, the probability for an exchange interaction of a
given binary is never much larger then $\sim0.1$, and usually much
smaller, with an exchange probability of $p_{\rm ex}\approx0.03$ being
typical. The probability is in particular much smaller than what one
would expect from an ``$nv\sigma$'' argument (or from the pre-factor
in equation [\ref{e:gx}]). The reason for the discrepancy is that binaries
typically do not survive for a relaxation time in the Galactic center,
but are prematurely disrupted by ionizations (even hard binaries can
move rapidly to higher energies, where they become soft), or in the
loss-cone.

\section{Summary and discussion}\label{s:disc}

In this paper we analyzed the evolution of single mass binary stars in a fixed environment of single stars and a MBH. We followed the evolution of the semi-major axis of the binaries, the energy and angular momentum with respect to the MBH, and the probability that a binary has experienced and exchange interaction. Channels for disruption of binaries are the ionization by a three body interaction, or tidal disruption by the MBH. The main goal was to provide a scheme to study the resulting dynamics, which we analyzed using a MC code.

We have compared simulations with $\Mbh=3\times10^6\Mo$ and $\Mbh=10^3\Mo$. Many of the properties of these systems are very similar. However, the mass of the MBH discriminates the systems in two ways. First, the tidal radius depends on $\Mbh$, and hence the loss-cone for binaries of a given hardness $\xi$ is different for the two systems. This is quantitatively expressed in equation (\ref{e:aemptyfull}). Second, assuming the relation between MBH mass and velocity dispersion of the host galaxy (equation \ref{e:Msigma}) implies that binaries of given hardness $\xi$ need to be tighter as $\Mbh$ increases. Since stars have finite radii, this implies that the lower the mass of the MBH, the harder the binaries can be, see equation (\ref{e:ah}).

It is perhaps worthwhile to stress several of the limitations of the approach we present in this paper, without attempting to provide an exhaustive list. 

The choice of assuming single mass stars obviously limits the use of our work, and it is of interest to consider a spectrum of masses. This is considerably more complicated, not only since the outcomes of the three-body interactions will then have much more parameters, but also since the dynamics with respect to the MBH will be mass dependent. In the current analysis we have also neglected dynamical friction by making the assumption that $D_y=0$ (\S\ref{s:DEE}). Integrating the multi-mass Fokker-Planck equations given in \citet{Bah77}, we find that for a two-mass population with the massive stars being twice as heavy and ten times as rare, the distribution of the light stars is not affected and has a $n\propto r^{-1.75}$ profile, while the distribution of the heavy stars (representing binaries) is slightly steeper with $n\propto r^{-1.9}$. This confirms that the distribution of the single stars is not significantly modified by the presence of the binaries, while the binaries themselves are driven towards the MBH by dynamical friction, although the effect is quite mild. On a qualitative level, this implies that the life time of binaries within the stellar cusp is further restricted, and they would be ionized more quickly (see also the discussion of the exchange probability below). 

Similarly, the inclusion of binary-binary interactions would make the analysis much harder. For both these extensions of our analysis, we believe that the study presented here will capture much of the dynamics and will therefore be useful to compare with. This is especially true for binary-binary interactions, which are very rare due to the small ($\sim 1-10\%$) binary fractions.

Another limitation is that we only evolved the stars bound to the MBH. Since the loss-cone is empty for all but the tightest binaries for the GC model, most of the tidal disruptions come in fact from orbits that are not bound to the MBH \citep[see, e.g.,][]{YuQ03, Per07}. The actual tidal disruption rate is thus much higher than we find in our analysis. It would be of interest to extend the treatment to consider unbound orbits; this can be done within the same MC scheme, see \citet{Mar79, Mar80}. As we noted in \S\ref{s:results}, it is in particular important to include global dynamics for IMBH systems, which have short relaxation times and cannot exist for a Hubble time in the state proposed here. 

A final limitation we stress here is that we did not include resonant relaxation \citep[e.g., ][]{Rau96, Rau98, Hop06, Gur07, Eil08, Per09b}. Throughout this paper, we have argued that loss-cone effects cannot dramatically change the DF, since energy diffusion operates on a comparable (somewhat shorter) time-scale. It was shown by \citet{Hop06} that resonant relaxation can change this. This relaxation mechanism is effective close to MBHs where the potential is close to a Kepler potential, so that orbits precess very slowly, leading to strong torques between the orbits. Since the resonant relaxation time can be much smaller than the non-resonant relaxation time, close to MBHs, energy diffusion (which still acts on the non-resonant relaxation time) is too slow to replenish binaries that are tidally disrupted. We expect that resonant relaxation time will modify the results presented here in the region very close to the MBH ($\sim0.01\pc$ for $\Mbh=3\times10^6\Mo$). While the MC method can be easily adapted to include this effect, we did not incorporate it since the details of resonant relaxation are not yet fully understood, in spite of ongoing efforts. In particular, it is not well known how the mechanism depends on back-reaction effects \citep[known as resonant friction, see][]{Rau96}.

As outlined in the introduction, there is considerable astrophysical motivation to study binaries in the vicinity of MBHs. We now consider a few of the ramifications that could be of observational relevance here.

{\it X-ray binaries in the Galactic center --- } There is an over-abundance of X-ray binaries in the Galactic center\footnote{we restrict our discussion to the inner pc of the Galactic center. At larger distances not only does our method not apply, but the relaxation time exceeds the Hubble time, implying that there will be very little dynamical evolution.} compared to the bulge \citep{Mun05}. Since there is very little evolution of the internal semi-major axis of the binary (see \S\ref{ss:rates} and \S\ref{ss:bin}), hardening cannot lead to additional X-ray binaries. Similarly, since the binary fraction drops within the radius of influence (see \S\ref{ss:fb}), the migration of X-ray binaries from the bulge to the central parsec can also not account for the relatively high number of X-ray binaries per unit stellar mass.

\citet{Mun05} suggested that the over-abundance of X-ray sources is
the result of exchange interactions. These can turn neutron stars into binary members, which can then continue to produce X-ray
binaries. We now compare their analysis to
the one made here. The mean probability for an exchange interaction of
a binary in the GC is $\langle
p_{\rm ex}\rangle=0.03$. Assuming that a fraction $f_{\rm NS}=0.01$ of
all stars in the GC are NSs, we find that the number of NSs that were
born as single objects, and are members of binaries at the end of the
simulation can be estimated as

\begin{equation}\label{e:bNS}
N_{\rm ex,\, NS} \approx f_{\rm NS}f_{b, 0}\int d^3r
\langle p_{\rm ex}\rangle f_b(r)n_s(r)\approx 20{f_{\rm NS}\over 0.01}{f_{b, 0}\over 0.05}.
\end{equation}

The number of binaries containing an exchanged compact remnant is
lower by a factor 10-100 than estimated in \citet{Mun05}, and about
25\% of all binaries should then burst every year in order to explain
the observed transient X-ray sources. The number of binaries
containing a NS can also be higher if the fraction of NS $f_{\rm NS}$
is higher than assumed here, or if the fraction of unbound hard
binaries is larger. A possible third cause of the relatively low event rate
estimated here is our simplifying assumption that all stars have
identical mass. Since compact remnants are likely to be more massive
than typical main sequence stars, they are possibly more likely to be
exchanged with one of the binary components than we assumed here. At
the same time, more massive binaries will drift towards the MBH more
rapidly, increasing the probability for ionization. It is not a priori obvious that accounting for differences in mass is actually conducive to the exchange rate of compact remnants. A treatment of binaries with a spectrum of masses is outside the
scope of this paper, but could be studied with a method similar to the
one presented here in future research.

{\it Hyper-velocity stars from the Large Magellanic Cloud --- }

For a MBH of mass similar to that in the Galactic center, the rate of tidal disruptions is dominated by binaries that are unbound to the MBH. This is not the case for hard binaries around IMBHs; as can be seen from figure (\ref{f:disr}d), the loss-cone becomes full at energies $E/v_0^2>1$. We can therefore use our simulations to estimate the tidal disruption rate near IMBHs.

Tidal disruptions of tight binaries lead to the production of hyper-velocity stars \citep{Hil88, YuQ03}. Most of the observed stars are consistent with being ejected from the Galactic center, but the star HE0437-5439 appears to be too young to have originated there \citep{Ede05}. This has led to speculations that this star has instead been ejected from an IMBH in the Large Magellanic Cloud \citep[LMC; see, e.g., ][]{Ede05, Gua05}, an hypothesis that appears to be supported by its metallicity \citep{Bon08, Pry08}. An alternative idea is that HE0437-5439 is a blue straggler \citep[e.g. ][]{Per07b}.

\citet{Gua05} performed three-body simulations to find the fraction of tidal disruptions by an IMBH that leads to the production of a hyper-velocity star with the properties of HE0437-5439. They found that for $a\approx0.1$ AU, a fraction $f^{\rm HVS}=0.1 f_{0.1}^{\rm HVS}$ of the tidal disruptions lead to the required velocities. From figure (\ref{f:disr}d), we estimate that the disruption rate per binary for such ratios is $N_0^{-1}d^2N/d\ln Edt\sim\,0.1\,{\rm Myr^{-1}}$. This result depends on the boundary conditions that are assumed, see \S\ref{ss:disr}. The boundary conditions here assume that IMBHs follow the relation between $\Mbh$ and $v_0$ (see equation \ref{e:Msigma}), consistent with evidence from $N$-body simulations for runaway mergers in young clusters \citep{Por02b}. Let $f^{b}=0.1f_{0.1}^{b}$ be the fraction of binaries with $a<0.1$ AU, and $f^{m>8}=0.01f_{0.01}^{m>8}$ the fraction of stars with mass $>8\Mo$, then the total number of relevant binaries is $N_0=10^3f^{b}f^{m>8}=f_{0.1}^{b}f_{0.01}^{m>8}$. We thus find that the rate at which an IMBH in the LMC could eject hypervelocity stars of the type of HE0437-5439 is $d^2(N/N_0)/d\ln Edt f^{\rm HVS} N_0 = 0.01\,{\rm Myr^{-1}}f_{0.1}^{b}f_{0.01}^{m>8}f_{0.1}^{\rm HVS}$. Assuming an age of $10\,\Myr$ for the age of the star, and accounting for a geometric factor such that $25\%$ of the ejected stars move towards the Galaxy, we find that the probability that we could observe a star like HE0437-5439 that originated in the LMC can be estimated as $p_{\rm LMC}=0.03f_{0.1}^{b}f_{0.03}^{m>8}f_{0.1}^{\rm HVS}$. This event rate is higher than that implied in \citet{Gua05}; see \citet{Per07b} for a discussion of how their ejection rates can be converted into detection rates. We conclude that the hypothesis of an origin in the LMC cannot be strongly rejected, although at the same time such an origin appears not to be very likely.

\acknowledgements{I thank Atakan G\"urkan, Douglas Heggie, Yuri Levin, and Hagai Perets and for
very stimulating discussions and advise, and the referee for a very helpful report. This work was supported by a Veni fellowship from the Netherlands Organization for Scientific Research (NWO). }

\appendix
\section[]{The  Fokker-Planck approximation}\label{s:appA}

At several places in this paper, we have used Fokker-Planck approximations, and MC methods to solve these equations. In this appendix we summarize several \citep[well known, see e.g.][]{Cha44} aspects of the Fokker-Planck approximation, and derive that these equations indeed describe the distribution of MC simulations.

Consider a general quantity $x$ of a star, which changes due to
dynamical interactions (this $x$ has no relation to earlier use of $x$ in the paper). If $\Psi(x, \Delta x)$ is the probability for
a change of size $\Delta x$ of a star at $x$, then the rate at which
the number of stars $n(x)dx$ in the range $(x, x+dx)$ changes is given
by the {\it master equation}

\begin{equation}\label{e:master}
{\partial n(x, t)\over\partial t} = \int d\Delta x n(x-\Delta x,
t)\Psi(x - \Delta x, \Delta x) - n(x, t)\int d\Delta x \Psi(x, \Delta x).
\end{equation}
The first term of the master equation gives the rate at which stars
flow into the bin $(x, x+dx)$, and the second term the rate at which
they flow out of this bin. The probability distribution $\Psi(x,
\Delta x)$ is determined by the ``micro physics'', such as two body
interactions, or interactions of a binary with a single star.

The master equation is an integro-differential equation. In many
situations in physics, the behavior of a system is determined mainly
by the sum of many small changes in a quantity, rather than rare but
strong encounters. In such situations, it is a good approximation to
make a second order Taylor approximation of the master equation
(\ref{e:master}). Approximating

\begin{equation}\label{e:Taylor}
\Psi(x - \Delta x, \Delta x)n(x-\Delta x, t) \approx \Psi(x, \Delta x)n(x,
t) - \Delta x{\partial \over\partial x} \left[\Psi(x, \Delta x)n(x,
t)\right] + {1\over 2}\Delta x^2{\partial^2 \over\partial x^2}
\left[\Psi(x, \Delta x)n(x, t)\right],
\end{equation}
and defining

\begin{equation}\label{e:diff}
D_{i}(x) \equiv \int d\Delta x  \Psi(x, \Delta x) \Delta^i,
\end{equation}
the {\it Fokker-Planck equation} can be written as

\begin{equation}\label{e:FP}
{\partial n(x, t)\over\partial t}= - {\partial \over \partial x}\left[ D_{1}(x) n(x, t)\right] + {1\over2}{\partial^2 \over \partial x^2}\left[ D_{2}(x) n(x, t)\right].
\end{equation}

From equation (\ref{e:diff}) it is clear that the diffusion
coefficients are just the moments of the scatter probability $\Psi$.

The first term in the Fokker-Planck equation clearly implies a drift
of stars in $x$. In a MC simulation, for a time-step $dt$
there is a deterministic step of size $D_{1}(x)dt$.

The second term represents a step in $x$ of size $\sqrt{ D_2(x) dt}$,
but with a random sign (with equal probability to be positive and
negative). To show this, we derive the diffusion equation for an
un-biased one dimensional random walk with varying step size.

Let the step-size at each time interval $\Delta t$ be $\Delta x$,
where $\Delta x=\Delta x(x)$ is a function of the quantity $x$ of the
star. First calculate the flux $\mathcal{F}(x, t)$ of stars through
$x$ at time $t$, i.e., the net rate at which stars flow from positions
smaller than $x$ to positions larger than $x$. To second order in
$\Delta x$, the smallest position from which a star can still flow
through $x$ is then

\begin{equation}
\Delta_{-}\equiv \Delta x(x-\Delta x) \approx \Delta x(x) - {d\Delta x\over
dx}\Delta x(x) \equiv \Delta x - {d\Delta x\over dx}\Delta x.
\end{equation}

Similarly, the largest position from which particles can step over $x$
is

\begin{equation}
\Delta_{+}\equiv \Delta x(x+\Delta x) \approx \Delta x(x) + {d\Delta x\over
dx}\Delta x(x) \equiv \Delta x + {d\Delta x\over dx}\Delta x.
\end{equation}

For an unbiased random walk, half of the particles step right, and
half of the particles step left. If at time $t$ the number of
particles in the interval $(x, x+dx)$ is given by $n(x, t)dx$, the
flux is then

\begin{equation}
\mathcal{F}(x, t) = {1\over2\Delta t}\int_{\Delta_{-}}^x dx n(x, t) - {1\over2\Delta t}\int_{x}^{\Delta_{+}}dx n(x, t).
\end{equation}

Using $\int_{x}^{x+h}dxf(x)\approx
hf(x+h/2)\approx f(x)h + {h^2\over2}{df\over
dx}$, the flux equals up to second order to

\begin{eqnarray}
\mathcal{F}(x, t) &=& {1\over2\Delta t}n(x-{\Delta\over2}, t)\Delta_{-} - {1\over2\Delta t}n(x+{\Delta\over2}, t)\Delta_{+} \nonumber\\
&=& {1\over2\Delta t}\left[ n(x, t) - {1\over2}{\partial n\over \partial x}\Delta x\right]\left[\Delta x - {d\Delta x\over dx}\Delta x\right] \nonumber\\
&&  - {1\over2 \Delta t}\left[ n(x, t) - {1\over2}{\partial n\over \partial x}\Delta x\right]\left[\Delta x - {d\Delta x\over dx}\Delta x\right] \nonumber\\
&=& -{1\over2}{\partial \over \partial x}\left[ {\Delta x^2\over \Delta t} n(x, t)\right]
\end{eqnarray}

Defining the diffusion coefficient

\begin{equation}
D_{2}(x)\equiv {\left[\Delta x(x)\right]^2\over \Delta t},
\end{equation}
and using the continuity equation 

\begin{equation}\label{e:cont}
{\partial n(x, t)\over \partial t}=-{\partial \over \partial x}\mathcal{F}(x, t),
\end{equation}
this leads to the diffusion equation

\begin{equation}
{\partial n(x, t)\over \partial t}= {1\over 2}{\partial^2\over \partial x^2}\left[ D_{2}(x)n(x, t)\right].
\end{equation}

Since this is the second term in the Fokker-Planck equation
(\ref{e:FP}), it follows that the Fokker-Planck equation can be solved
by a MC method in which stars make at each time step a
deterministic step of size $D_{1}(x)dt$ and a step of magnitude
$\sqrt{ D_2(x) dt}$ and random sign.


\end{document}